\documentclass[ALICE,manyauthors]{cernphprep}
\usepackage[comma,square,numbers,sort&compress]{natbib}

\usepackage{lineno}
\usepackage{xspace}
\usepackage{hyperref}
\usepackage[T1]{fontenc}
\usepackage{orcidlink}

\begin{document}
%

\newcommand{\pp}           {pp\xspace}
\newcommand{\ppbar}        {\mbox{$\mathrm {p\overline{p}}$}\xspace}
\newcommand{\XeXe}         {\mbox{Xe--Xe}\xspace}
\newcommand{\PbPb}         {\mbox{Pb--Pb}\xspace}
\newcommand{\pA}           {\mbox{pA}\xspace}
\newcommand{\pPb}          {\mbox{p--Pb}\xspace}
\newcommand{\AuAu}         {\mbox{Au--Au}\xspace}
\newcommand{\dAu}          {\mbox{d--Au}\xspace}

\newcommand{\s}            {\ensuremath{\sqrt{s}}\xspace}
\newcommand{\snn}          {\ensuremath{\sqrt{s_{\mathrm{NN}}}}\xspace}
\newcommand{\pt}           {\ensuremath{p_{\rm T}}\xspace}
\newcommand{\meanpt}       {$\langle p_{\mathrm{T}}\rangle$\xspace}
\newcommand{\ycms}         {\ensuremath{y_{\rm CMS}}\xspace}
\newcommand{\ylab}         {\ensuremath{y_{\rm lab}}\xspace}
\newcommand{\etarange}[1]  {\mbox{$\left | \eta \right |~<~#1$}}
\newcommand{\yrange}[1]    {\mbox{$\left | y \right |~<~#1$}}
\newcommand{\dndy}         {\ensuremath{\mathrm{d}N_\mathrm{ch}/\mathrm{d}y}\xspace}
\newcommand{\dndeta}       {\ensuremath{\mathrm{d}N_\mathrm{ch}/\mathrm{d}\eta}\xspace}
\newcommand{\avdndeta}     {\ensuremath{\langle\dndeta\rangle}\xspace}
\newcommand{\dNdy}         {\ensuremath{\mathrm{d}N_\mathrm{ch}/\mathrm{d}y}\xspace}
\newcommand{\Npart}        {\ensuremath{N_\mathrm{part}}\xspace}
\newcommand{\Ncoll}        {\ensuremath{N_\mathrm{coll}}\xspace}
\newcommand{\dEdx}         {\ensuremath{\textrm{d}E/\textrm{d}x}\xspace}
\newcommand{\RpPb}         {\ensuremath{R_{\rm pPb}}\xspace}

\newcommand{\nineH}        {$\sqrt{s}~=~0.9$~Te\kern-.1emV\xspace}
\newcommand{\seven}        {$\sqrt{s}~=~7$~Te\kern-.1emV\xspace}
\newcommand{\twoH}         {$\sqrt{s}~=~0.2$~Te\kern-.1emV\xspace}
\newcommand{\twosevensix}  {$\sqrt{s}~=~2.76$~Te\kern-.1emV\xspace}
\newcommand{\five}         {$\sqrt{s}~=~5.02$~Te\kern-.1emV\xspace}
\newcommand{\twosevensixnn}{$\sqrt{s_{\mathrm{NN}}}~=~2.76$~Te\kern-.1emV\xspace}
\newcommand{\fivenn}       {$\sqrt{s_{\mathrm{NN}}}~=~5.02$~Te\kern-.1emV\xspace}
\newcommand{\LT}           {L{\'e}vy-Tsallis\xspace}
\newcommand{\GeVc}         {Ge\kern-.1emV/$c$\xspace}
\newcommand{\MeVc}         {Me\kern-.1emV/$c$\xspace}
\newcommand{\TeV}          {Te\kern-.1emV\xspace}
\newcommand{\GeV}          {Ge\kern-.1emV\xspace}
\newcommand{\MeV}          {Me\kern-.1emV\xspace}
\newcommand{\GeVmass}      {Ge\kern-.2emV/$c^2$\xspace}
\newcommand{\MeVmass}      {Me\kern-.2emV/$c^2$\xspace}
\newcommand{\lumi}         {\ensuremath{\mathcal{L}}\xspace}

\newcommand{\ITS}          {\rm{ITS}\xspace}
\newcommand{\TOF}          {\rm{TOF}\xspace}
\newcommand{\ZDC}          {\rm{ZDC}\xspace}
\newcommand{\ZDCs}         {\rm{ZDCs}\xspace}
\newcommand{\ZNA}          {\rm{ZNA}\xspace}
\newcommand{\ZNC}          {\rm{ZNC}\xspace}
\newcommand{\SPD}          {\rm{SPD}\xspace}
\newcommand{\SDD}          {\rm{SDD}\xspace}
\newcommand{\SSD}          {\rm{SSD}\xspace}
\newcommand{\TPC}          {\rm{TPC}\xspace}
\newcommand{\TRD}          {\rm{TRD}\xspace}
\newcommand{\VZERO}        {\rm{V0}\xspace}
\newcommand{\VZEROA}       {\rm{V0A}\xspace}
\newcommand{\VZEROC}       {\rm{V0C}\xspace}
\newcommand{\Vdecay} 	   {\ensuremath{V^{0}}\xspace}

\newcommand{\ee}           {\ensuremath{e^{+}e^{-}}} 
\newcommand{\pip}          {\ensuremath{\pi^{+}}\xspace}
\newcommand{\pim}          {\ensuremath{\pi^{-}}\xspace}
\newcommand{\kap}          {\ensuremath{\rm{K}^{+}}\xspace}
\newcommand{\kam}          {\ensuremath{\rm{K}^{-}}\xspace}
\newcommand{\pbar}         {\ensuremath{\rm\overline{p}}\xspace}
\newcommand{\kzero}        {\ensuremath{{\rm K}^{0}_{\rm{S}}}\xspace}
\newcommand{\lmb}          {\ensuremath{\Lambda}\xspace}
\newcommand{\almb}         {\ensuremath{\overline{\Lambda}}\xspace}
\newcommand{\Om}           {\ensuremath{\Omega^-}\xspace}
\newcommand{\Mo}           {\ensuremath{\overline{\Omega}^+}\xspace}
\newcommand{\X}            {\ensuremath{\Xi^-}\xspace}
\newcommand{\Ix}           {\ensuremath{\overline{\Xi}^+}\xspace}
\newcommand{\Xis}          {\ensuremath{\Xi^{\pm}}\xspace}
\newcommand{\Oms}          {\ensuremath{\Omega^{\pm}}\xspace}
\newcommand{\degree}       {\ensuremath{^{\rm o}}\xspace}

\newcommand{\figref}[1]{Fig.~\ref{#1}}
\newcommand{\Figref}[1]{Figure~\ref{#1}}
\newcommand{\secref}[1]{Sec.~\ref{#1}}
\newcommand{\Secref}[1]{Section~\ref{#1}}

\begin{titlepage}
\PHyear{2022}       
\PHnumber{111}      
\PHdate{01 June}  

\title{Closing in on critical net-baryon fluctuations at LHC energies:\\ cumulants up to third order in Pb--Pb collisions}
\ShortTitle{Net-baryon fluctuations with cumulants up to third order in Pb--Pb collisions}   

\Collaboration{ALICE Collaboration\thanks{See Appendix~\ref{app:collab} for the list of collaboration members}}
\ShortAuthor{ALICE Collaboration} 

\begin{abstract}
Fluctuation measurements are important sources of information on the mechanism of particle production at LHC energies. This article reports the first experimental results on third-order cumulants of the net-proton distributions in Pb--Pb collisions at a center-of-mass energy \mbox{$\snn=5.02$~TeV} recorded by the ALICE detector. The results on the second-order cumulants of net-proton distributions at $\snn = 2.76$ and $5.02$~TeV are also discussed in view of effects due to the global and local baryon number conservation. The results demonstrate the presence of long-range rapidity correlations between protons and antiprotons. Such correlations originate from the early phase of the collision. The experimental results are compared with HIJING and EPOS model calculations, and the dependence of the fluctuation measurements on the phase-space coverage is examined in the context of lattice quantum chromodynamics (LQCD) and hadron resonance gas (HRG) model estimations. The measured third-order cumulants are consistent with zero within experimental uncertainties of about 4\% and are described well by LQCD and HRG predictions.
\end{abstract}
\end{titlepage}

\setcounter{page}{2} 


\section{Introduction}  \label{Intro}

Predictions based on the theory of the strong interaction, QCD, imply that, at sufficiently high energy densities, nuclear matter transforms into a state called quark--gluon plasma (QGP)~\cite{Shuryak:1980tp,Ding:2015ona}, where chiral symmetry is restored and quarks and gluons are deconfined. Ultrarelativistic heavy-ion collisions are ideal environments to study the phase diagram of strongly interacting matter and the physics of the QGP state as a function of temperature ($T$) and baryon chemical potential ($\mu_{\rm B}$). While the QCD phase diagram is largely unknown for $\mu_{\rm B} > 450$ MeV, the region below that value down to $\mu_{\rm B} = 0$ has been well explored theoretically~\cite{Ding:2015ona,Borsanyi:2013bia} and experimentally~\cite{Andronic:2017pug}. In that region, the chiral phase transition is most likely a continuous crossover with pseudo-critical temperature $T_{\mathrm{pc}} \approx 157$~MeV at $\mu_{\rm B} = 0$~\cite{HotQCD:2018pds, Borsanyi:2020fev}. Near $\mu_{\rm B} = 0$ the properties of the QCD phase transition depend on the number of quark flavors and their masses. For vanishing masses of the light quarks (u,d) the transition is of second order and belongs to the universality class of three-dimensional O(4)-symmetric spin models~\cite{Pisarski:1983ms}. The small u,d quark masses of order of 1\% of the constituent quark masses constitute a small but explicit breaking of chiral symmetry. For these small physical quark masses LQCD indicates that the transition turns into a smooth crossover. To date, there is no experimental confirmation of the crossover nature of the transition. Nevertheless, the smallness of the physical quark masses may leave traces of critical behavior also for a crossover transition. Therefore, a significant effort at the RHIC and LHC colliders is concentrated on quantifying the nature of the phase transition in the small $\mu_{\rm B}$ region.

For large $\mu_{\rm B}$ the phase diagram may exhibit a ``critical endpoint (CEP)". The search for the CEP is one of the main physics goals of the beam energy scan programs at RHIC ~\cite{Luo:2017faz,Xu:2018vnf}, at the CERN SPS ~\cite{NA49:2009diu}, and at FAIR~\cite{CBM:2016kpk}. 

Recent LQCD~\cite{HotQCD:2019xnw,Borsanyi:2018grb,Bazavov:2011nk} calculations of chiral susceptibilities, derivatives of the chiral condensate with respect to quark masses, exhibit for small quark masses a clear peak at $T_{\mathrm{pc}}$, consistent with the chemical freeze-out temperature extracted by the analysis of hadron multiplicities~\cite{Andronic:2017pug,ALICE:2013jfw} measured in central Pb--Pb collisions by the ALICE experiment. This suggests that the chemical freeze-out occurs very close to the chiral phase transition at LHC energies. Critical signals associated with this phenomenon can be linked with long-range correlations and increasing multiplicity fluctuations due to the existence of the massless modes of the second-order phase transition~\cite{Friman:2011pf,ParisenToldin:2003hq}. In particular, the fluctuations of the conserved charges are very sensitive probes for the equation of state and can be directly related to the thermodynamic susceptibilities, which are calculable in the framework of LQCD. The quark-number susceptibilities are defined as the derivatives of the reduced QCD pressure ($P/T^{4}$) with respect to the reduced chemical potentials ($\hat{\mu}=\mu/T$) of the conserved charges

\begin{equation}
    \label{eq:mixed_cumulant}
\chi_{klmn}^{\mathrm{B,S,Q,C}}=\frac{\partial^{(k+l+m+n)} (P(\hat{\mu}_{\rm B},\hat{\mu}_{\rm S},\hat{\mu}_{\rm Q},\hat{\mu}_{\rm C})/T^{4})}{\partial \hat{\mu}_{\rm B}^{k}\partial \hat{\mu}_{\rm S}^{l}\partial \hat{\mu}_{\rm Q}^{m}\partial \hat{\mu}_{\rm C}^{n}    }\Big|_{\vec{\mu}=0}.
\end{equation}

Here the relevant conserved charges, represented by the chemical potentials, are the electric charge Q, the baryon number B, the strangeness S, and the charm C. These susceptibilities are studied experimentally in terms of the ratios of the cumulants\footnote{The cumulants, $\kappa_{n}$, of net-baryon number, $\Delta N_{\rm B}=N_{\rm B}-N_{\rm \overline{B}}$, are defined as the coefficients in the Maclaurin series of the logarithm of the characteristic function of $\Delta N_{\rm B}$~\cite{Braun-Munzinger:2016yjz}.} of net-charge distributions~\cite{Gavai:2010zn}. Here and in the following, net-charge distribution stands for the difference between the distributions of positively and negatively charged particles, where ``charge" refers to any additive quantum number. Signs of criticality due to the proximity of the chiral crossover transition to a second-order transition are expected to show up starting only with the sixth-order cumulants of net-charge distributions~\cite{Ejiri:2005wq,Friman:2011pf}. 
Measuring conserved charge fluctuations in ultrarelativistic nuclear collisions is a challenging task, see Ref.~\cite{Luo:2017faz} for a recent review. In addition to possible critical fluctuations due to the proximity of a CEP or O(N) criticality, there are several other dynamical signals, such as correlations due to baryon number conservation~\cite{Bzdak:2012an,Braun-Munzinger:2019yxj,Vovchenko:2020kwg}, volume fluctuations~\cite{Braun-Munzinger:2016yjz}, thermal blurring~\cite{Ohnishi:2016bdf}, resonance decays~\cite{Bluhm:2020mpc,Vovchenko:2020kwg}, initial-state fluctuations~\cite{Shen:2017bsr}, baryon annihilation~\cite{Savchuk:2021aog}, ``excluded volume" effects~\cite{Vovchenko:2020kwg}, etc., which could overshadow the dynamical fluctuations of interest.

The article is organized as follows. In \Secref{Data}, details are given about the ALICE detector, the data set and the analysis procedure, such as event and track selection criteria, particle identification and the efficiency correction technique. In \Secref{Results}, results on second-order cumulants of net-pion and net-kaon distributions are presented to demonstrate the influence of resonance decays on the measured cumulants.  The main results are contained in the second- and third-order cumulants of net-proton distributions and compared to theoretical expectations. The article concludes, in \Secref{Summary}, with a discussion on how the present results fit into the general strategy to get information on possible critical behavior near the QCD phase boundary and with an outline of the next steps.

\section{Experimental setup and data analysis} \label{Data}
 
The measurements presented in this article are based on about 13 and 78 million minimum bias Pb--Pb collisions at $\snn = 2.76$ and $5.02$~TeV recorded with the ALICE detector~\cite{ ALICE:2008ngc, ALICE:2014sbx} in the years 2010 and 2015, respectively. Note that the running conditions have changed significantly in 2015. 
The details presented in this article refer only to the analysis of 2015 data (see Refs.~\cite{ ALICE:2019nbs, ALICE:2017jsh} for 2010). The minimum-bias trigger condition is defined by requiring a coincidence of hits in both \VZERO detectors~\cite{ALICE:2013axi} located on either side of the nominal interaction point along the beam direction and covering the pseudorapidity intervals \mbox{$2.8 < \eta < 5.1$} and \mbox{$-3.7 < \eta < -1.7$}. The definition of the centrality~\cite{ ALICE:2013hur} is based on the signal amplitudes measured in the \VZERO detectors, which are related to the collision geometry and the number of participating nucleons through a Monte Carlo (MC) simulation based on a Glauber model~\cite{Loizides:2014vua,Miller:2007ri}. Beyond event characterization, the main sub-detectors used in the analysis are the Time Projection Chamber (\TPC)~\cite{Alme:2010ke} for tracking and particle identification (PID), using the specific energy loss \dEdx, and the Inner Tracking System (ITS)~\cite{CERN-LHCC-99-012,ALICE:2010tia} for tracking and vertex determination.

Event and track selection criteria were applied to ensure optimal PID performance via \dEdx measurement and momentum ($p$) resolution, as well as good track quality. The \VZERO~\cite{ALICE:2013axi} and the Zero Degree Calorimeter~\cite{ALICE:1999edx} timing information was used to reject background due to beam--gas interactions and parasitic beam--beam interactions. Along the beam direction, events occurring within $\pm0.15$~cm, where tracking is affected by the central membrane of the TPC, and outside $\pm7$~cm of the nominal interaction point were discarded to keep detection efficiency uniform as a function of event vertex position. This corresponds to 24\% of the total events.

The analysis uses tracks of charged particles reconstructed using the \ITS and the TPC in the pseudorapidity range of $|\eta|< 0.8$ and with full azimuthal acceptance. Due to the different gas mixture, higher interaction rates, and larger multiplicities in 2015, the TPC performance was affected by local space-charge distortions caused by the accumulation of space charge originating from the gaps between adjacent readout chambers~\cite{Arslandok:2022dyb}. Although these distortions are corrected on average during reconstruction, their fluctuations caused a significant mismatch between the TPC and ITS tracks. They affected all events detected within a time interval of about 0.5 seconds after a collision, corresponding to the full drift time of ions from the amplification region in the TPC readout chambers to the central electrode. Collisions within such time intervals were discarded using the event interaction time (timestamp), when in addition  the mean matching efficiency between the TPC and ITS for all tracks was less than 88\%. The nominal TPC--ITS track matching efficiency is narrowly peaked at 90.5\%. By applying this condition about 7\% of all events are rejected. Tracks were accepted if their $\chi^{2}$ value per space point from the track fit is less than 3.5. To improve the \dEdx resolution and reduce the contributions from track splitting\footnote{Track splitting occurs when a track is reconstructed as two separate tracks due to poor \dEdx resolution.}, two additional track selection criteria were applied. The number of clusters used for the \dEdx calculation\footnote{Clusters that are very close to the TPC readout chamber boundaries~\cite{Arslandok:2022dyb} or are from overlapping tracks are not considered in the calculation of \dEdx. Therefore, the number of clusters used to calculate \dEdx may differ from the number of clusters used for track selection. This quantity is particularly important for the \dEdx resolution.} was set to more than 70, and short tracks crossing less than 80 out of a maximum of 159 pad rows were excluded from the analysis. These criteria address both tracks crossing chamber boundaries and split tracks, both leading to a deterioration of the \dEdx measurement. To suppress contributions from particles from weak decays and interactions in the detector material, the distance of closest approach (DCA) of the extrapolated track to the primary vertex position was required to be less than 1~cm along the beam direction, while in the transverse plane a transverse momentum (\pt) dependent DCA selection of less than (0.018 $\rm cm$ $+$ 0.035$\pt^{-1.01}$) with $\pt$ in GeV/\textit{c} was applied to account for the \pt dependence of the DCA resolution~\cite{ ALICE:2010yje}. In addition, daughter tracks from reconstructed secondary weak-decay kink topologies were discarded.

In 2015, the ALICE TPC was operated at interaction rates up to 8~kHz during Pb--Pb collisions. These high interaction rates resulted in a possible pile-up of several interactions during the 90 $\mu$s drift time. The fraction of pile-up events that occur within a few cm of the primary vertex, the so-called in-bunch pile-up, is less than 1\textperthousand\ of the entire data set. Therefore, in-bunch pile-up has a negligible impact on the results. Moreover, in-bunch pile-up events were further suppressed thanks to the high-precision vertexing capabilities of the ITS. Even slightly displaced multiple vertices originating from interactions within the same bunch crossing can be discriminated. The much more frequent out-of-bunch pile-up events occur anytime during the drift time of the TPC, i.e., they are distributed equally along the time direction within the TPC and cannot be suppressed by reconstructed secondary vertices in the ITS. They lead to a significant deterioration in the measurements of the \dEdx of the triggered event caused by the baseline fluctuations in the TPC~\cite{Arslandok:2022dyb}. These pile-up events affect about 20\% of the data collected in 2015 and are negligible for 2010. The resulting bias in the measured \dEdx is corrected to restore optimal PID performance and to avoid the need to discard events with pile-up~\cite{Arslandok:2022dyb}.

Protons are identified by their \dEdx in the TPC and its known momentum dependence. To overcome the misidentification problem caused by overlapping \dEdx distributions for different particles, for instance at the crossings of kaons (K) and protons (p) or pions ($\pi$) and protons, a novel experimental technique, the Identity Method (IM)~\cite{Gazdzicki:2011xz,Rustamov:2012bx,Arslandok:2018pcu} was used. With this method, weights, which are obtained from the fits of \dEdx distributions, are assigned to each track reflecting the probability of a particle having a specific identity. Thereby all tracks can be kept without applying any selection on the PID variable and no second detector needs to be employed to identify the (anti)protons. In the present analysis, particles in the momentum range 0.6--2.0~GeV/\textit{c} were retained. With particle identification based solely on the \dEdx measurements in the TPC, the efficiencies are independent of momentum in the selected range, and as high as about 91\% for protons and about 83\% for antiprotons. The difference of 8\% is due to the absorption of antiprotons in the detector material. Both proton and antiproton detection efficiencies are nearly independent of collision centrality and are uniform within the kinematic acceptance used in this analysis. The cumulants of the net-proton distribution are reconstructed using the IM as well. Further details on the application of the IM to ALICE data are discussed in Ref.~\cite{ ALICE:2017jsh}. In Ref.~\cite{ ALICE:2019nbs}, the method used here was applied to the analysis of second-order cumulants of net protons in Pb--Pb collisions at $\snn = 2.76$~TeV. The current analysis closely follows what is described there. 

\begin{figure}[htbp]
  \centering
     \begin{minipage}{.5\textwidth}
         \vspace*{\fill}
         \centering
         	\includegraphics[width=\textwidth]{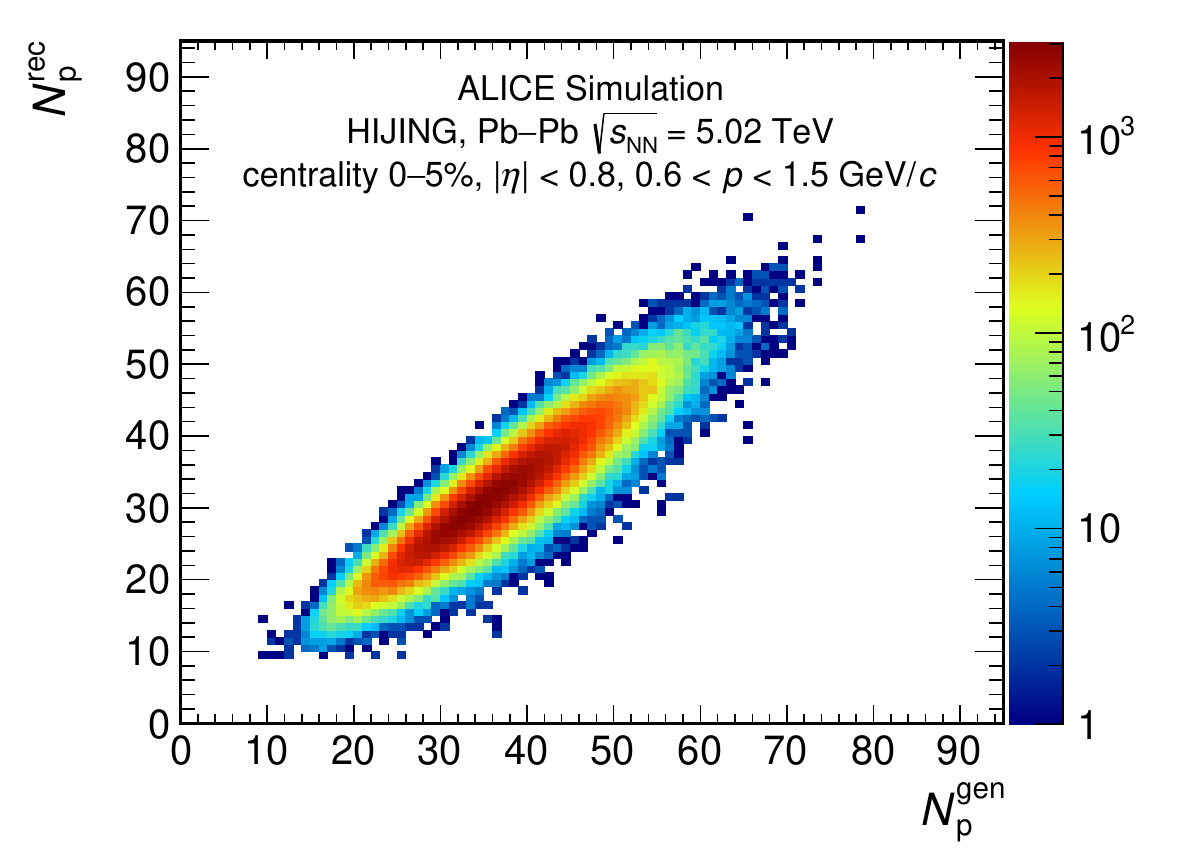}
     \end{minipage}%
     \begin{minipage}{.5\textwidth}
         \vspace*{\fill}
         \centering
         \includegraphics[width=\textwidth]{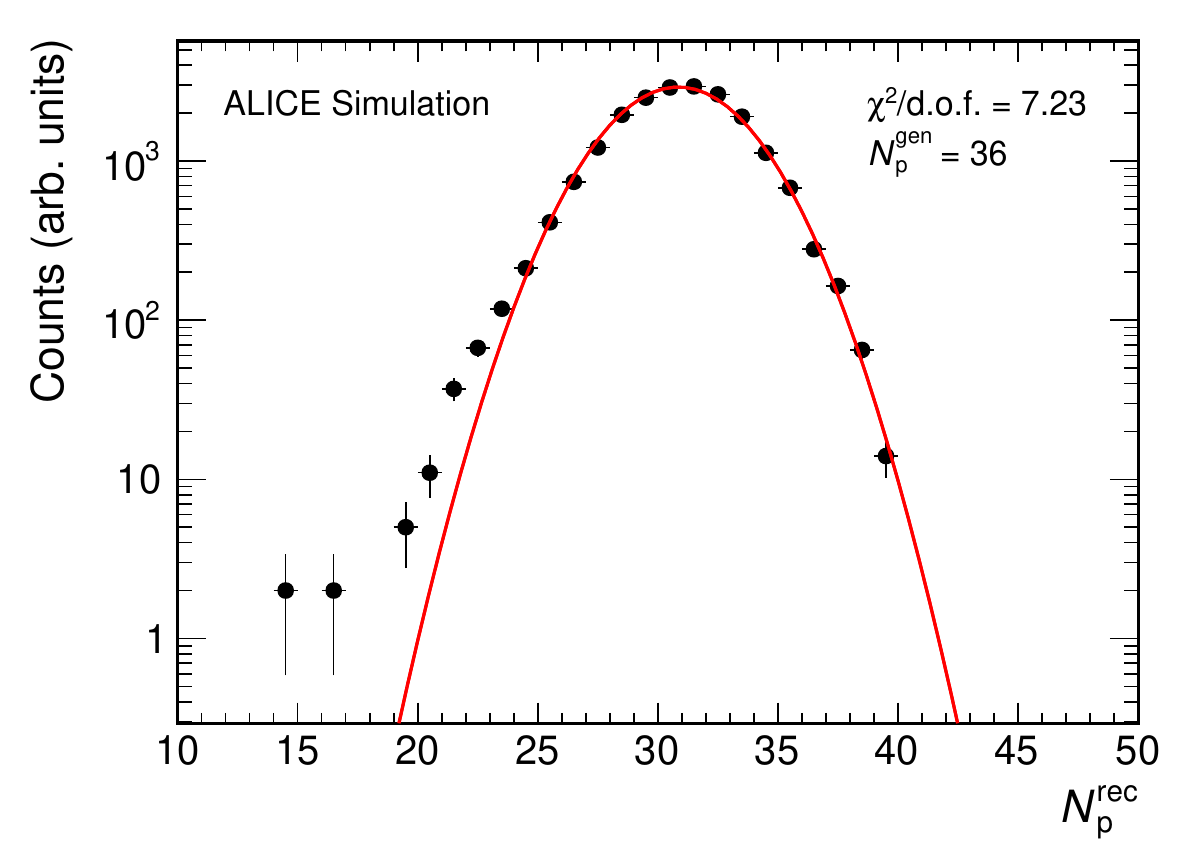}
     \end{minipage}%
  \caption{(Left) Correlation between the reconstructed ($N_{\rm p}^{\rm rec}$) and the generated ($N_{\rm p}^{\rm gen}$) number of protons for the most central Pb--Pb collisions simulated using the HIJING model~\cite{Gyulassy:1994ew}. (Right) Distribution of reconstructed proton number for a fixed value of $N_{\rm p}^{\rm gen}=36$, where the fit demonstrates the deviation from a binomial efficiency loss.}
  \label{fig:2D_det_response}
\end{figure}

Understanding and controlling the particle detection efficiency is one of the major technical challenges in the measurement of higher-order cumulants, since the efficiency enters into the analytical formula of the correction with the corresponding higher power~\cite{Bzdak:2012ab, Bzdak:2013pha, Nonaka:2017kko}. This affects both statistical and systematic uncertainties of the corrected data. Note that the efficiency correction approximately doubles the statistical uncertainties, as also noted in Ref.~\cite{Pandav:2018bdx}. The efficiency correction for the cumulants is performed by using proton and antiproton efficiencies in analytic formulas derived in Refs.~\cite{Bzdak:2012ab, Bzdak:2013pha, Nonaka:2017kko} assuming efficiency losses governed by the binomial statistics (binomial efficiency loss). In order to ensure that this assumption is fulfilled, the ALICE detector response~\cite{ALICE:2014sbx} was studied in detail with a full MC simulation using the HIJING event generator and the GEANT4~\cite{GEANT4:2002zbu} transport software. The response of the TPC detector for the kinematic range used in this analysis is illustrated in the left panel of \figref{fig:2D_det_response} as a correlation between the reconstructed ($N_{\rm p}^{\rm rec}$) and the generated ($N_{\rm p}^{\rm gen}$) number of protons, where the reconstructed protons are those detected in the active area of the detector, as well as satisfying the event and track quality criteria. The event and track selection plays a significant role in the shape of the TPC detector response. For example, as discussed above, they can reduce the contributions from track splitting, which causes correlated \dEdx measurements, and thus a deviation from a binomial detector response. 

\begin{figure}[htbp]
  \centering
     \begin{minipage}{.5\textwidth}
         \vspace*{\fill}
         \centering
         	\includegraphics[width=\textwidth]{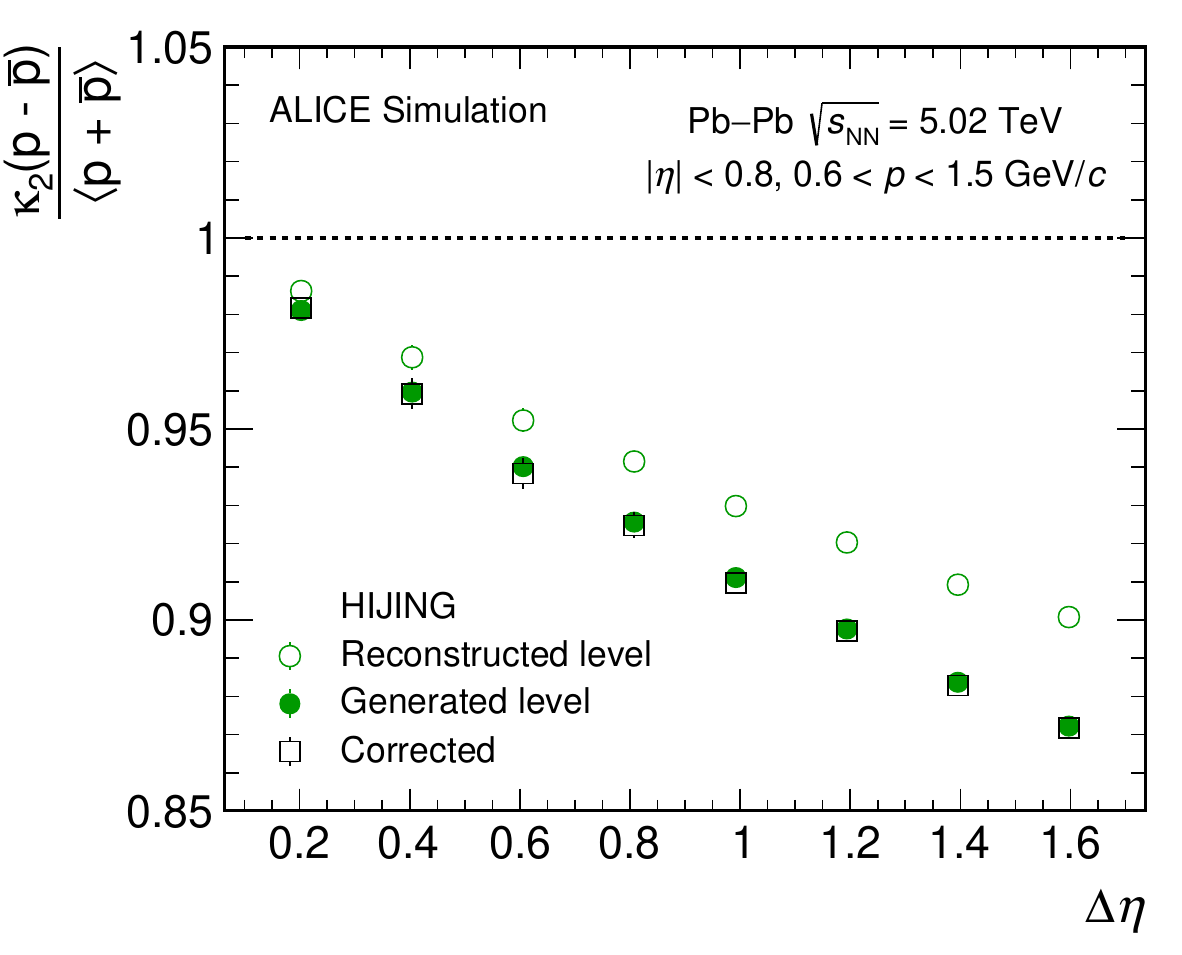}
     \end{minipage}%
     \begin{minipage}{.5\textwidth}
         \vspace*{\fill}
         \centering
         \includegraphics[width=\textwidth]{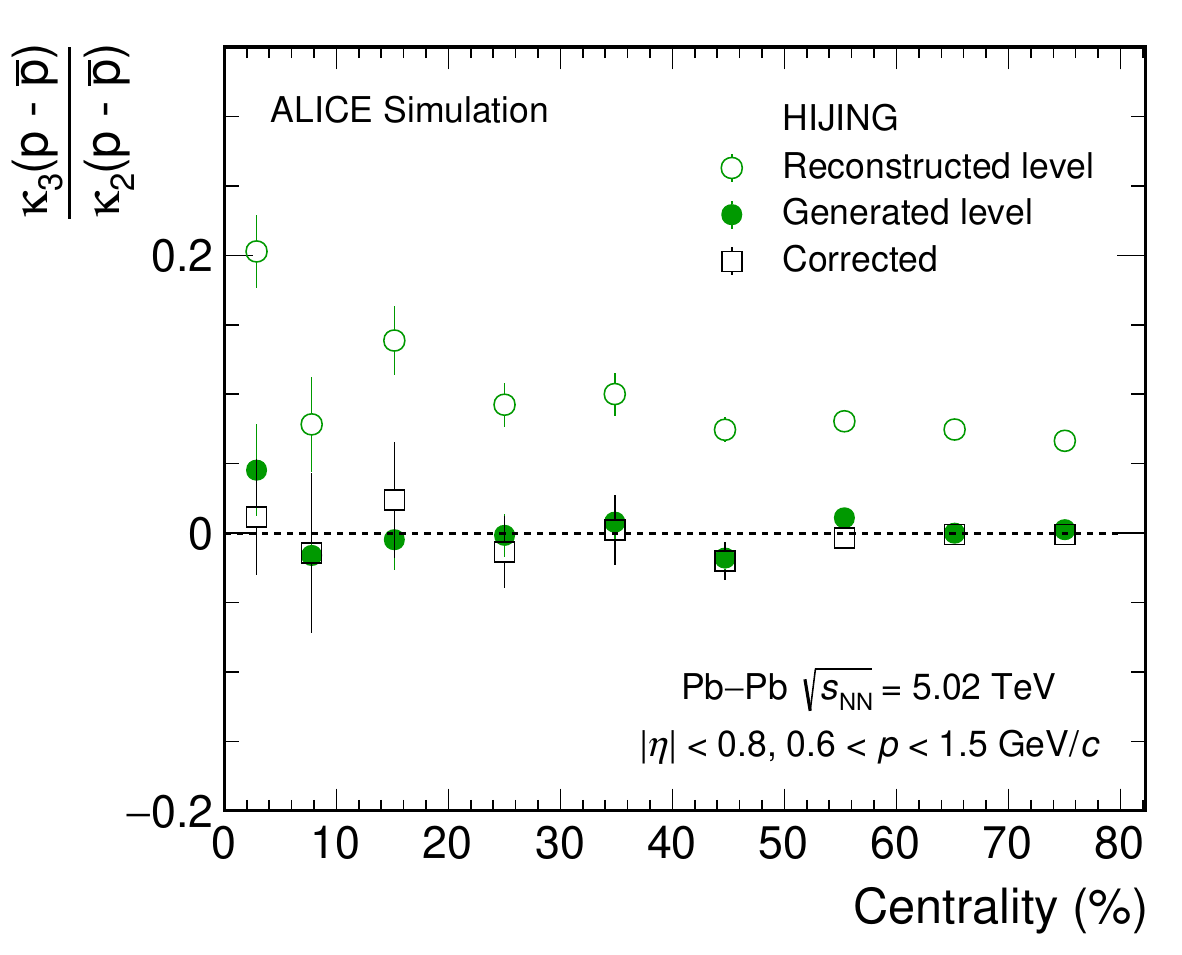}
     \end{minipage}%
  \caption{HIJING model~\cite{Gyulassy:1994ew} based calculations of the normalized second-order cumulants of net protons as a function of pseudorapidity window ($\Delta\eta$) (left) and ratio of third- to second-order cumulants (right) of net protons as a function of collision centrality at $\snn=5.02$~TeV. The results at the generated and reconstructed level are shown by the green closed and open circles, respectively. The error bars represent statistical uncertainties. The results after efficiency correction assuming binomial efficiency losses~\cite{Bzdak:2012ab, Bzdak:2013pha, Nonaka:2017kko} are shown by black open squares.}
  \label{fig:MC_closure_C2OverSkellam_eta}
\end{figure}

After detailed study and optimization of the event and track selection, only a slight deviation from the binomial loss is observed, as shown in the right panel of \figref{fig:2D_det_response}. A MC verification test was performed to estimate the impact of this deviation in the final results on the net-proton cumulant measurements. In this MC closure test, particles are generated, including certain correlations such as the effect of baryon number conservation, and reconstructed after they have passed through the detector simulated with GEANT4. Then the efficiency correction is applied, and the generated and corrected observables are compared. The comparison is shown in \figref{fig:MC_closure_C2OverSkellam_eta} for the second- and third-order cumulant ratios of the net-proton distribution. The efficiency-corrected results obtained from the MC reconstructed data are in agreement with the results obtained from the MC generated data. Note that the uncorrected results, also in \figref{fig:3rd_CentDep_Raw}, deviate from the Skellam baseline, which is set at zero, because the net-proton number is nonzero due to antiproton absorption in the detector material and, to a smaller degree, by a proton knock-out contribution. Therefore, the final results depend crucially on a very accurate determination of the proton and antiproton efficiencies.

The statistical uncertainties assigned to the reconstructed cumulants were determined using the sub-sample method. To this end, the data set was subdivided into $n$ random subsamples. The distribution of the cumulants from these subsamples yields the statistical uncertainty as described in Ref.~\cite{ ALICE:2017jsh}. The fits to the measured \dEdx distributions, which are the only inputs to the IM, are the dominant source of systematic uncertainty in the ratios of the cumulants, for both second and third order. The observed maximum deviation between fit variations~\cite{ALICE:2017jsh} is 0.6\% and 0.8\% for the normalized second-order cumulants within the momentum intervals of 0.6--1.5~GeV/\textit{c} and 0.6--2.0~GeV/\textit{c}, respectively, and 4\% for the ratio of third- to second-order cumulants in the momentum interval 0.6--1.5~GeV/\textit{c}. The impact of possible imperfections in the \dEdx correction procedure mentioned above is also included in this systematic uncertainty estimate by analyzing the data retaining different fractions of events containing pile-up. The uncertainties associated with the detection efficiencies of the (anti)protons are also investigated by varying the detection efficiencies by an amount of $\pm2$\% for protons and antiprotons separately. The resulting systematic variation is less than 0.2\% and 1.5\% for the second- and third-order cumulants, respectively. Other sources of systematic uncertainty are estimated by varying the event and track selection criteria, resulting in a maximum uncertainty of less than 1\%. The final total systematic uncertainty is obtained by adding in quadrature the individual maximum systematic deviations from these three groups of independent contributions. For the third-order cumulants, it varies between less than 0.5\% for the most peripheral collisions and a maximum of 3\% for the most central collisions for the pseudorapidity interval of $\Delta \eta=1.6$.

\section{Results} \label{Results}

As potential candidates for conservation of electric charge and strangeness, results are reported for the pseudorapidity interval dependence of the second-order cumulants of net-pions and net-kaons produced in central Pb--Pb collisions.

\begin{figure}[htbp]
  \centering
     \begin{minipage}{.48\textwidth}
         \vspace*{\fill}
         \centering
         	\includegraphics[width=\textwidth]{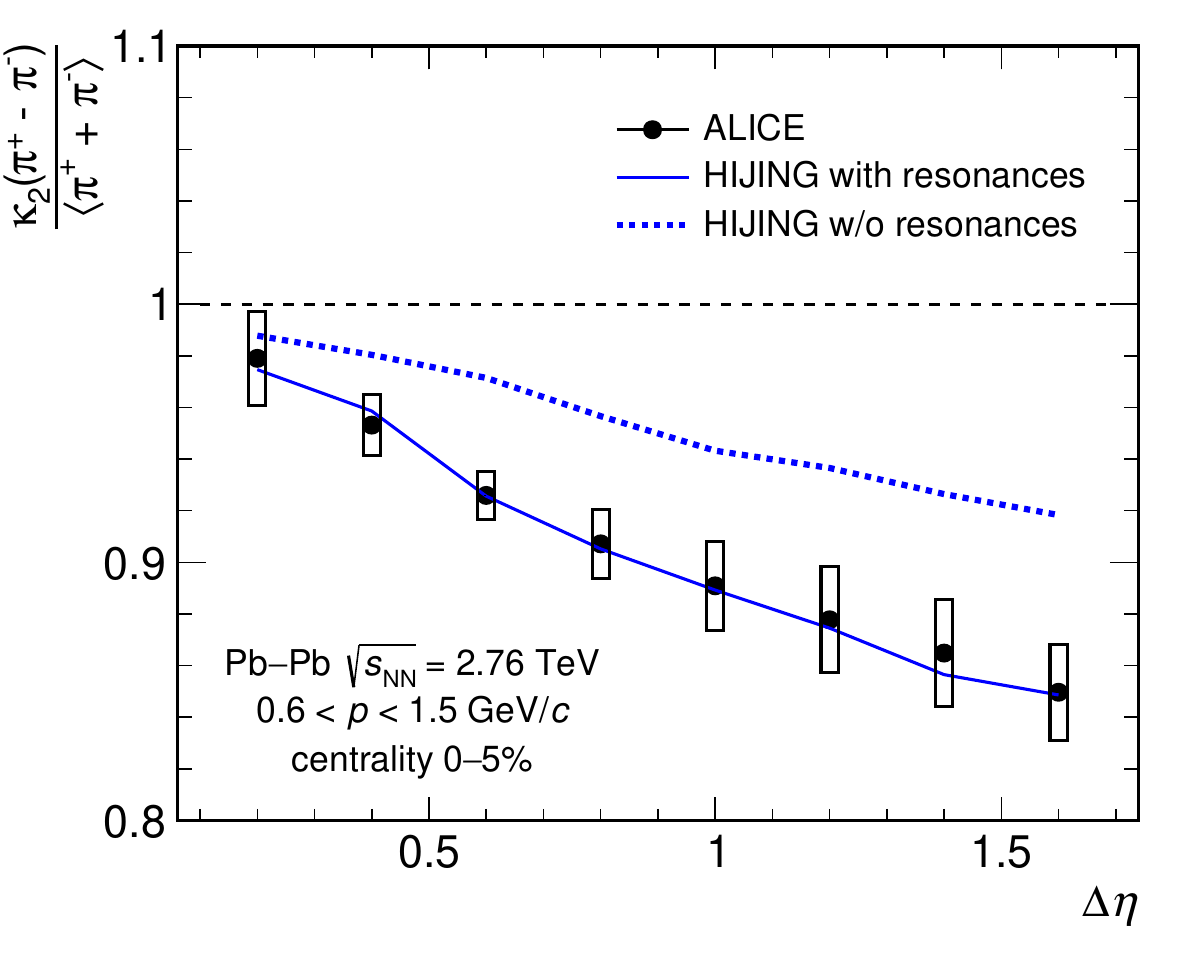}
     \end{minipage}%
     \hspace*{0.5cm}
     \begin{minipage}{.48\textwidth}
         \vspace*{\fill}
         \centering
         \includegraphics[width=\textwidth]{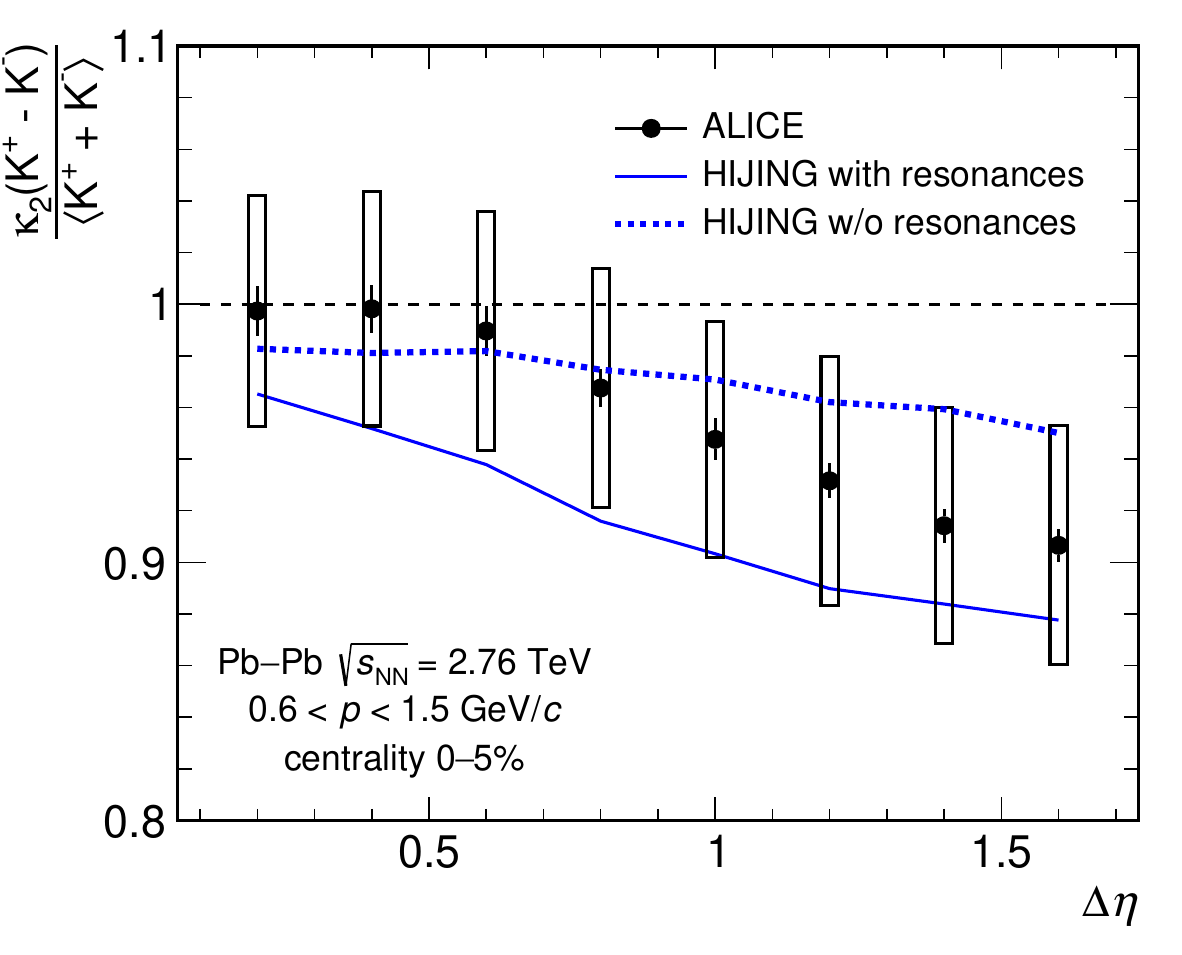}
\end{minipage}%
  \caption{Pseudorapidity interval dependence of the second-order cumulants of net-pions (left) and net-kaons (right) normalized to the means (see text). The ALICE data are shown as solid black circles while the blue solid and dashed lines indicate the results from HIJING~\cite{Gyulassy:1994ew} model calculations with and without resonance contributions, respectively. The error bars represent
  statistical uncertainties and the boxes around the data points represent the total systematic uncertainties.}
  \label{fig:net_pions_wo_resonances}
\end{figure}

The observations in these channels are quite striking because they shed light on resonance decay contributions to fluctuations in Pb--Pb collisions at the LHC. \Figref{fig:net_pions_wo_resonances} shows the pseudorapidity interval dependence of the normalized second-order cumulants of net-pions and net-kaons compared with the results from HIJING~\cite{Gyulassy:1994ew} with and without resonance contributions. A significant effect of resonances, e.g., $\rho \rightarrow \pi^{+}\pi^{-}$ and $\phi \rightarrow \rm K^{+}\rm K^{-}$, is clearly visible in both cases. In fact, the decay of resonances into oppositely charged pion or kaon pairs drastically reduces the fluctuations and dominates the second-order cumulants of the respective net distributions. Therefore, to study the genuine electric charge and strangeness fluctuations, first a quantitative understanding of the resonance contributions is essential. On the other hand, there are no resonances that decay into $\rm p\overline{\rm p}$ with a sizeable branching ratio, therefore net-proton fluctuations are not obscured by this effect. It has been argued in the literature~\cite{Kitazawa:2012at} that net-proton fluctuations are good proxies for net-baryon fluctuations, in particular for $\snn > $10~GeV. Also, total electric-charge conservation is expected to have a negligible impact on the net-proton fluctuation measurements, since the electric charge is mostly carried by the charged pions, which are the most abundant species at LHC energies. The statistically independent Poisson limit for net-baryon distributions is the Skellam distribution, which is defined as the probability distribution of the difference of two random variables, each generated from statistically independent Poisson distributions~\cite{Braun-Munzinger:2011xux,Skellam}. For net protons, the $n$th-order cumulants of the Skellam distribution are given by
\begin{equation}
    \label{eq:skellam}
\kappa^{\mathrm{Skellam}}_{n}(\mathrm{p}-\mathrm{\overline{p}}) = \langle \mathrm{p} \rangle + (-1)^{n} \langle \mathrm{\overline{p}} \rangle,
\end{equation}
where $\langle \mathrm{p} \rangle$ and $\langle \mathrm{\overline{p}} \rangle$ are the mean values of the proton and antiproton multiplicity distributions, respectively. That means that even-order cumulants of the Skellam distribution of the net protons are just the sum of the mean numbers of protons and antiprotons. At LHC energies, these numbers are equal within 1\%~\cite{ALICE:2013mez}, and therefore the normalized cumulants of the Skellam distribution with respect to its second-order cumulant are zero for odd cumulants and unity for even cumulants. At $T_{\mathrm{pc}}$~\cite{HotQCD:2018pds,Borsanyi:2020fev}, both the predictions based on LQCD and the HRG~\cite{Andronic:2017pug} model agree with the Skellam baseline up to the third-order cumulants of the net protons, reflecting independent Poissonian fluctuations. The LQCD prediction~\cite{Bazavov:2020bjn}, including the effect of dynamical quarks, shows a significant deviation from the Skellam baseline for the fourth- and higher-order cumulants, while the standard HRG does not contain such effects and deviations from the Skellam baseline are only due to baryon number conservation~\cite{Braun-Munzinger:2020jbk}. Fluctuations of conserved charges are meaningful only within a limited phase space. They vanish in the full phase space, in order to obey the conservation laws, and asymptotically approach the Poisson limit for very small acceptance, where dynamical correlations are suppressed~\cite{Bzdak:2012ab}. Therefore, the fluctuations of net-baryons are studied in the framework of the Grand Canonical Ensemble, where the net-baryon number is conserved only on average. Accordingly, the analysis is performed differentially as a function of the collision centrality, the pseudorapidity interval, $\Delta \eta=$~0.2 to 1.6, and for two different momentum ranges, 0.6--1.5~GeV/\textit{c} and 0.6--2.0~GeV/\textit{c}. It should be noted that the determination of centrality and the net-proton analysis are based on measurements in different pseudorapidity intervals to avoid trivial effects due to autocorrelations~\cite{Braun-Munzinger:2016yjz}.

\begin{figure}[htbp]
  \centering
     \begin{minipage}{.5\textwidth}
         \vspace*{\fill}
         \centering
         	\includegraphics[width=\textwidth]{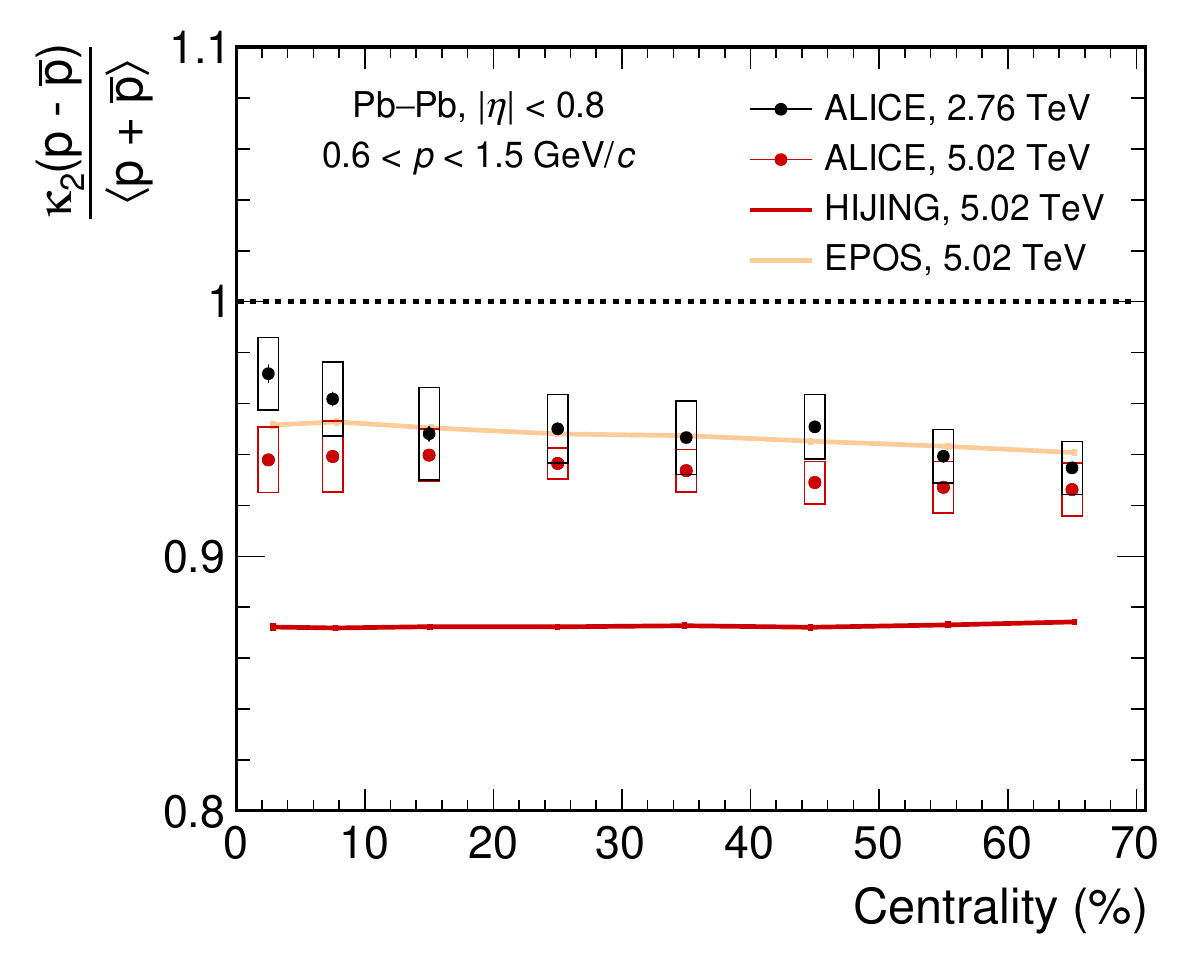}
     \end{minipage}%
     \begin{minipage}{.5\textwidth}
         \vspace*{\fill}
         \centering
         \includegraphics[width=\textwidth]{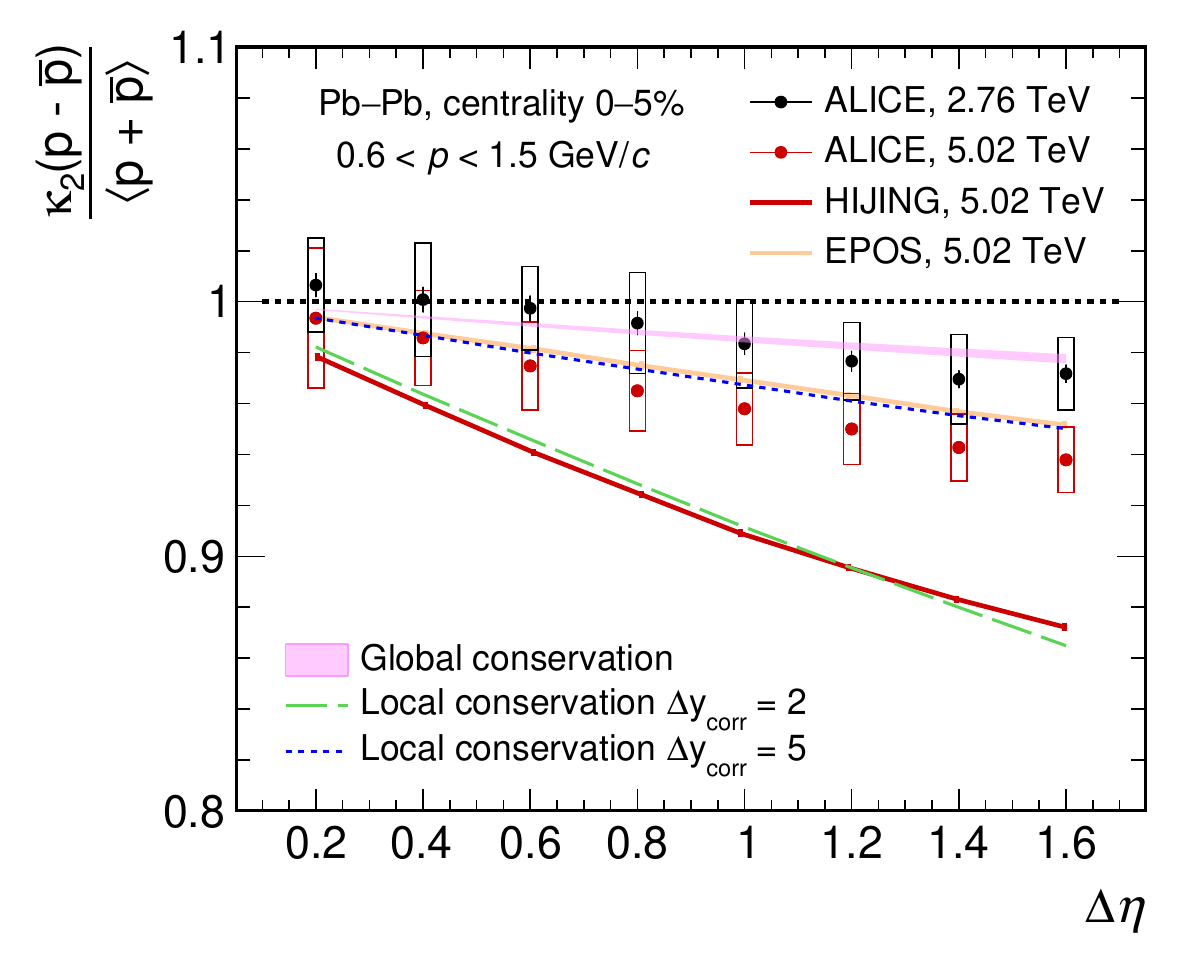}
     \end{minipage}%
  \caption{(Color online) Centrality (left) and pseudorapidity interval (right) dependence of the normalized second-order cumulants of net protons. The ALICE data are shown by black and red markers for $\snn = 2.76$ and $5.02$~TeV, respectively, while the colored shaded areas indicate the results from HIJING~\cite{Gyulassy:1994ew} and EPOS~\cite{Pierog:2013ria} model calculations at $\snn =5.02$~TeV. The Skellam baseline is shown by the horizontal dashed black line. In the right panel the expectation from global baryon number conservation is shown as a pink band and the dashed colored lines represent the predictions of the model with local baryon number conservation~\cite{Braun-Munzinger:2019yxj}. 
}
  \label{fig:2nd_CentDep_10hComp}
\end{figure}

\Figref{fig:2nd_CentDep_10hComp} shows the measured centrality and pseudorapidity dependence of the normalized second-order cumulants of the net protons in Pb--Pb collisions for the two collision energies. The 5.02 TeV data appear to be somewhat lower, however the two data sets agree within systematic uncertainties. It should be noted that the systematic uncertainties exhibit a large degree of correlation from bin to bin, but between the two collision energies are essentially uncorrelated due to the different running conditions (collision rate, gas mixture and space charge distortions in the TPC, etc.). The normalized second-order cumulants are independent of collision centrality and are reduced by about 5\% from the Skellam baseline for the $\Delta\eta=1.6$ interval (left panel). As a function of the width of the $\Delta\eta$ interval, the fluctuations are increasingly reduced. Due to the increasing relevance of baryon number conservation with larger acceptance, this is expected. For the narrowest interval, statistically independent Poissonian fluctuations are observed. The results are also compared to results from HIJING~\cite{Gyulassy:1994ew} and EPOS~\cite{Pierog:2013ria} model calculations at \mbox{$\snn = 5.02$~TeV}. HIJING treats nucleus--nucleus collisions as an independent superposition of nucleon--nucleon interactions and does not include phenomena such as equilibrium and collectivity. While in HIJING the hadronization is based on the Lund string fragmentation scheme, EPOS (version 1.99, tuned to LHC data) distinguishes between string segments in a collectively behaving central part (``core") with high energy density and those in a peripheral part (``corona") with lower energy density, more like in pp or p--A collisions.  

It is noteworthy that the proton to antiproton ratio is above unity in both models with a significance of more than 3 sigma: 1.025$\pm 0.004$ and 1.008$\pm 0.002$ for EPOS and HIJING, respectively; the value measured by ALICE in Pb--Pb collisions at $\snn=2.76$~TeV agrees with unity within experimental uncertainties of a few percent~\cite{ ALICE:2013mez}. This implies that the volume fluctuations for the second- and third-order cumulants are not negligible for the model calculations~\cite{Braun-Munzinger:2016yjz}.

Both model calculations show second-order fluctuations independent of centrality, as do the experimental data. The results from the EPOS calculations agree within the uncertainties with the data both in terms of the centrality and the $\Delta\eta$ dependencies. The HIJING model results exhibit a 12\% suppression compared to Poissonian fluctuations for the widest $\Delta\eta$ interval and are significantly below the data. This is also apparent in the dependence on the width of the $\Delta\eta$ interval (right panel of \figref{fig:2nd_CentDep_10hComp}). 
The dependence on acceptance, and specifically the discrepancy between the HIJING results and the data, is examined in view of global vs local baryon number conservation modelled in Refs.~\cite{Braun-Munzinger:2016yjz, ALICE:2019nbs,Braun-Munzinger:2019yxj,Braun-Munzinger:2020jbk} using a canonical statistical model. The right panel of \figref{fig:2nd_CentDep_10hComp} shows the results for different widths of the correlation interval, ranging from global baryon number conservation to short-range correlations. As expected, measurements and model calculations converge to the Skellam baseline in the limit of very small acceptance. As already noted in Ref.~\cite{ ALICE:2019nbs},
the data from ALICE indicate long-range rapidity correlations ($\Delta y_{\rm corr} > 5$) between protons and antiprotons, therefore originating from the early phase of the collision~\cite{Dumitru:2008wn}. Here $\Delta y_{\rm corr}/2$ is defined as the correlation length between protons and antiprotons in rapidity~\cite{Braun-Munzinger:2019yxj}, so that $\Delta y_{\rm corr}=5$ means that protons are correlated with antiprotons within 2.5 rapidity units into either direction.

\begin{figure}[htbp]
  \centering
     \begin{minipage}{.5\textwidth}
         \vspace*{\fill}
         \centering
         	\includegraphics[width=\textwidth]{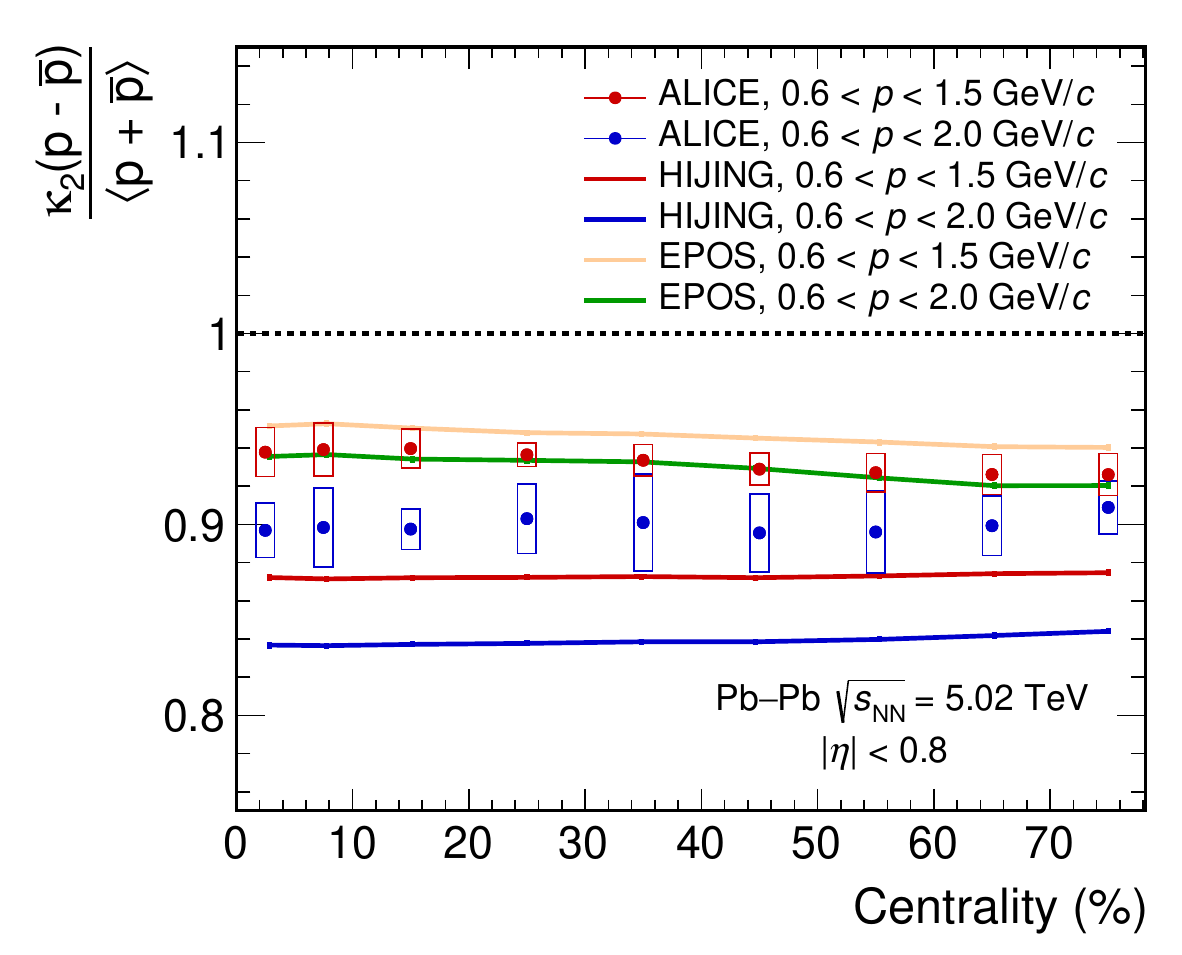}
     \end{minipage}%
     \begin{minipage}{.5\textwidth}
         \vspace*{\fill}
         \centering
         \includegraphics[width=\textwidth]{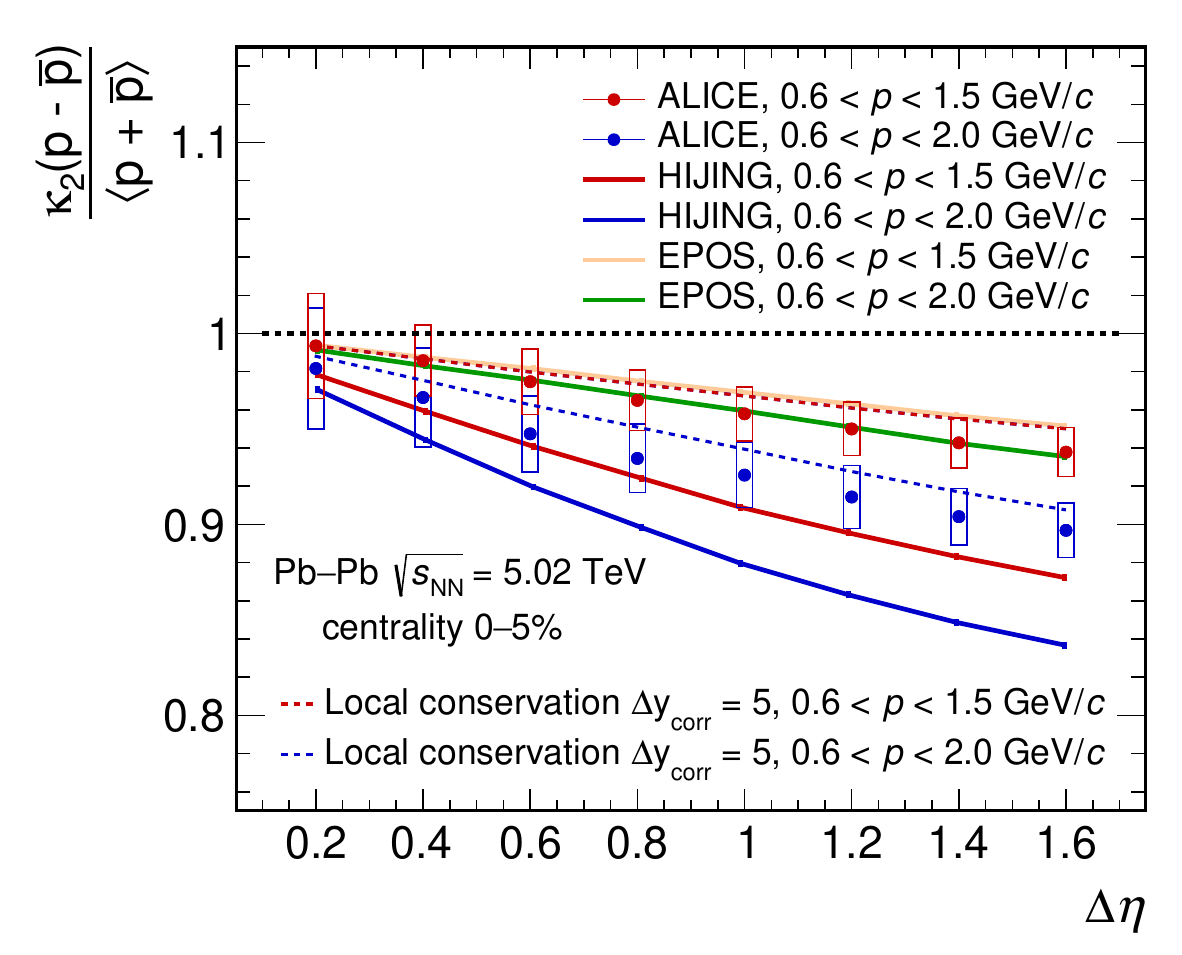}
     \end{minipage}%
  \caption{(Color online) Centrality (left) and pseudorapidity interval (right) dependence of the normalized second-order cumulants of net protons for \mbox{$\snn = 5.02$~TeV} and two momentum intervals for the protons. The ALICE data are shown by red and blue markers for $0.6 < p < 1.5$~GeV/\textit{c} and $0.6 < p < 2.0$~GeV/\textit{c}, respectively. The colored shaded areas indicate the results from the HIJING~\cite{Gyulassy:1994ew} and EPOS~\cite{Pierog:2013ria} model calculations. In the right panel, in addition, the dashed colored lines represent the predictions from the model with local baryon number conservation with $\Delta y_{\rm corr}=5$~\cite{Braun-Munzinger:2019yxj}.}
  \label{fig:2nd_CentDep}
\end{figure}

The HIJING model calculations reflect a much smaller correlation length (\mbox{$\Delta y_{\rm corr}=2$}) than the EPOS model (\mbox{$\Delta y_{\rm corr}=5$}) and the ALICE data. This is likely due to the formation of baryons in string breaking in the underlying Lund string model~\cite{Andersson:1983ia}. This sensitivity to the range of proton--antiproton correlations is further studied by enlarging the momentum acceptance for the \mbox{$\snn = 5.02$~TeV} data. The resulting normalized second-order cumulants are shown in \figref{fig:2nd_CentDep} for the momentum intervals 0.6--1.5~GeV/\textit{c} and 0.6--2.0~GeV/\textit{c}. Using the wider momentum interval the number of protons and antiprotons roughly doubles. While there is again no change with the collision centrality, the larger acceptance leads to a larger suppression of fluctuations for the wider momentum range. For the largest $\Delta\eta$ interval, the suppression amounts to an additional 4\%. All calculations reflect the reduction in the fluctuations. However, while the magnitude is properly reproduced by the canonical statistical model predictions with a long correlation length~\cite{Braun-Munzinger:2019yxj}, it can be noted that the suppression due to increased acceptance is somewhat weaker in the EPOS results. The HIJING calculations properly track the absolute reduction in fluctuations, but fall significantly below the data in absolute amount.

\begin{figure}[htbp]
  \centering
     \begin{minipage}{.5\textwidth}
         \vspace*{\fill}
         \centering
         	\includegraphics[width=\textwidth]{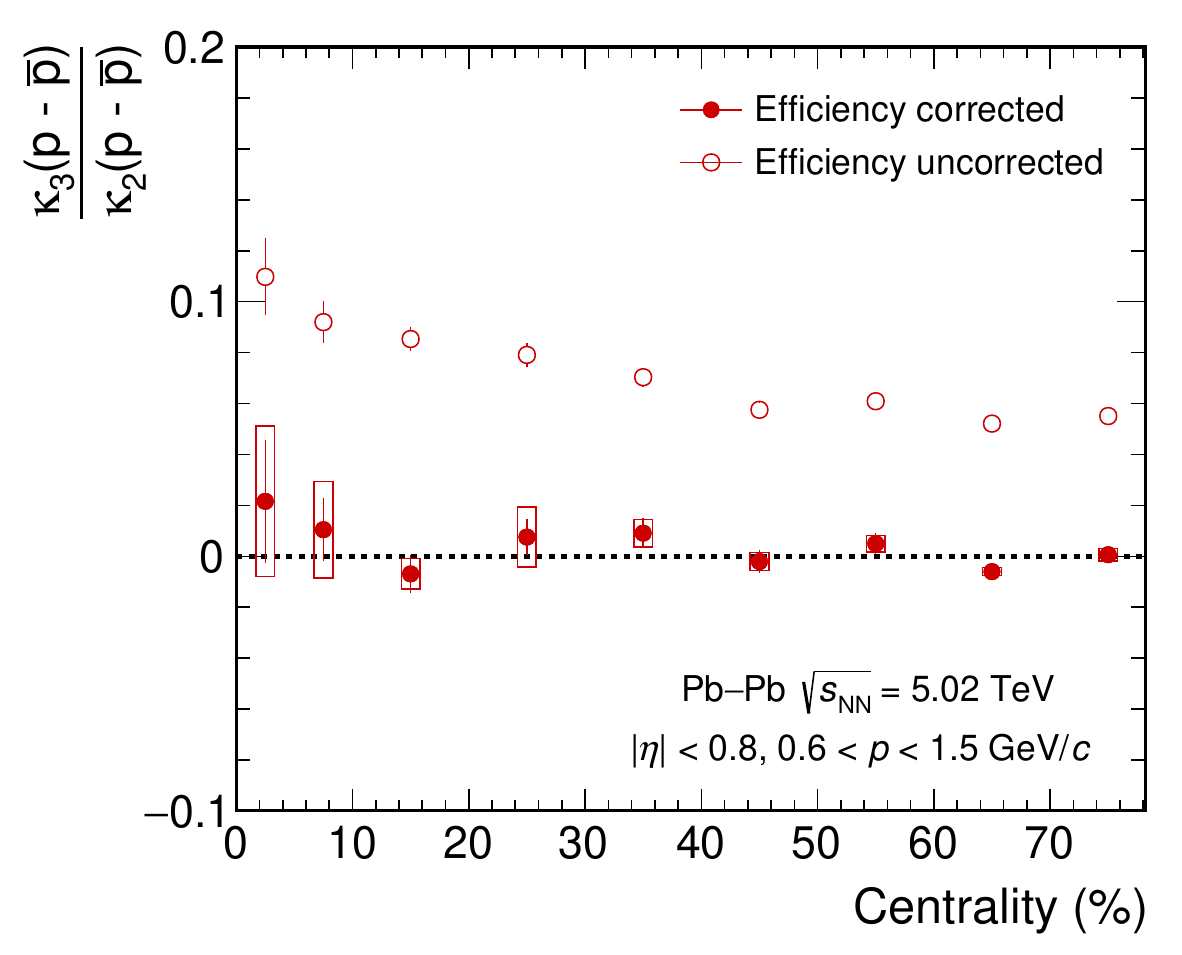}
     \end{minipage}%
     \begin{minipage}{.5\textwidth}
         \vspace*{\fill}
         \centering
         \includegraphics[width=\textwidth]{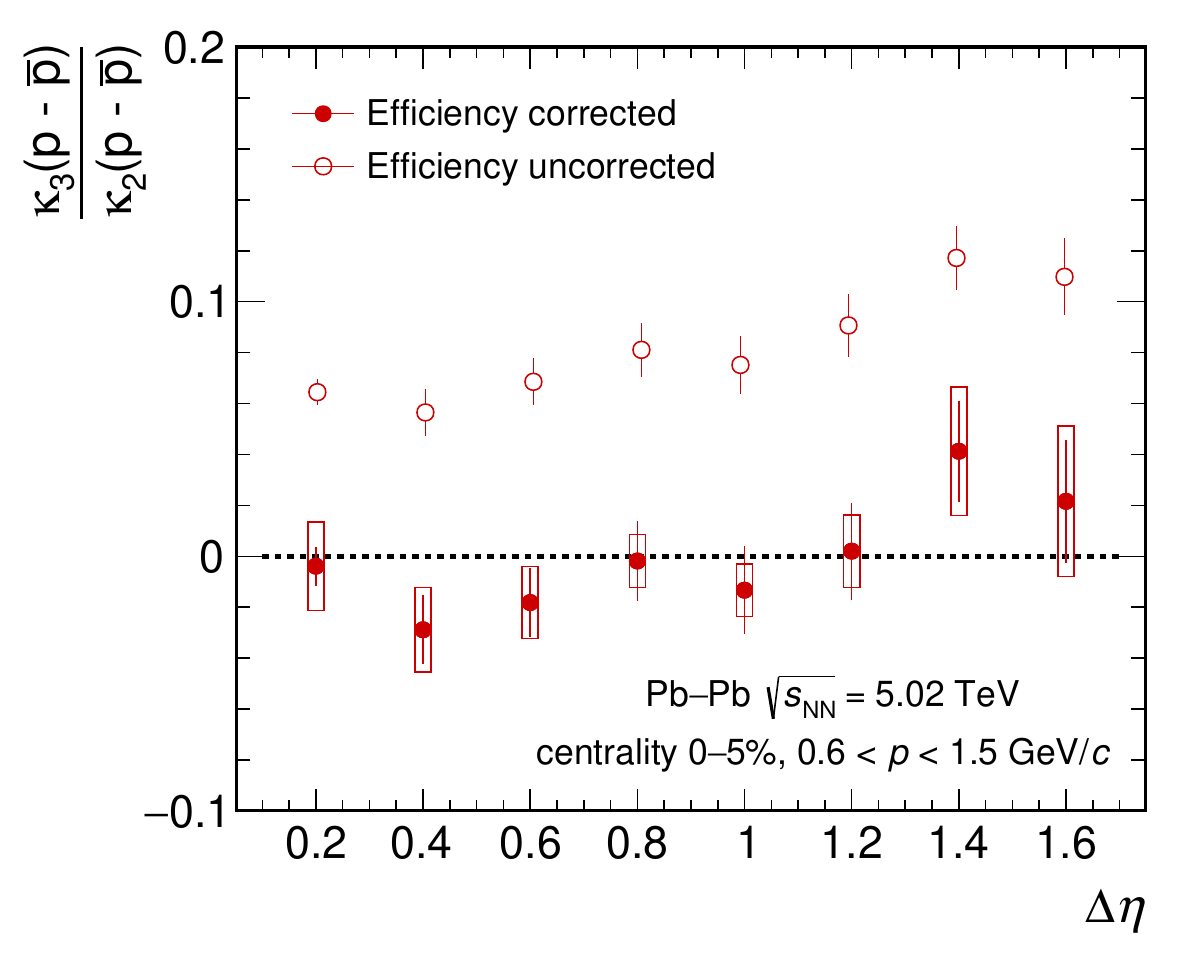}
     \end{minipage}%
  \caption{Centrality (left) and pseudorapidity interval (right) dependence of the ratio of third- to second-order cumulants for net protons at \mbox{$\snn = 5.02$~TeV} before (open markers) and after (closed markers) efficiency correction.}
  \label{fig:3rd_CentDep_Raw}
\end{figure}

\Figref{fig:3rd_CentDep_Raw} shows the centrality and pseudorapidity dependence of the ratio of third- to second-order cumulants of net protons at \mbox{$\snn = 5.02$~TeV} before and after the efficiency correction. After efficiency correction, the data agree with the zero baseline within the experimental uncertainties, which is consistent with expectations from the HRG model. The experimentally achieved overall precision is better than 4\% for the most central collisions and much smaller for more peripheral collisions. Note that in the HRG model all odd cumulants vanish at LHC energy, where the number of baryons and antibaryons agree. The odd cumulants vanish under these conditions also if baryon number conservation is included, see Refs.~\cite{Braun-Munzinger:2018yru,Braun-Munzinger:2020jbk}. Also in LQCD~\cite{HotQCD:2017qwq} the odd cumulants vanish. 

\begin{figure}[htbp]
  \centering
     \begin{minipage}{.5\textwidth}
         \vspace*{\fill}
         \centering
         	\includegraphics[width=\textwidth]{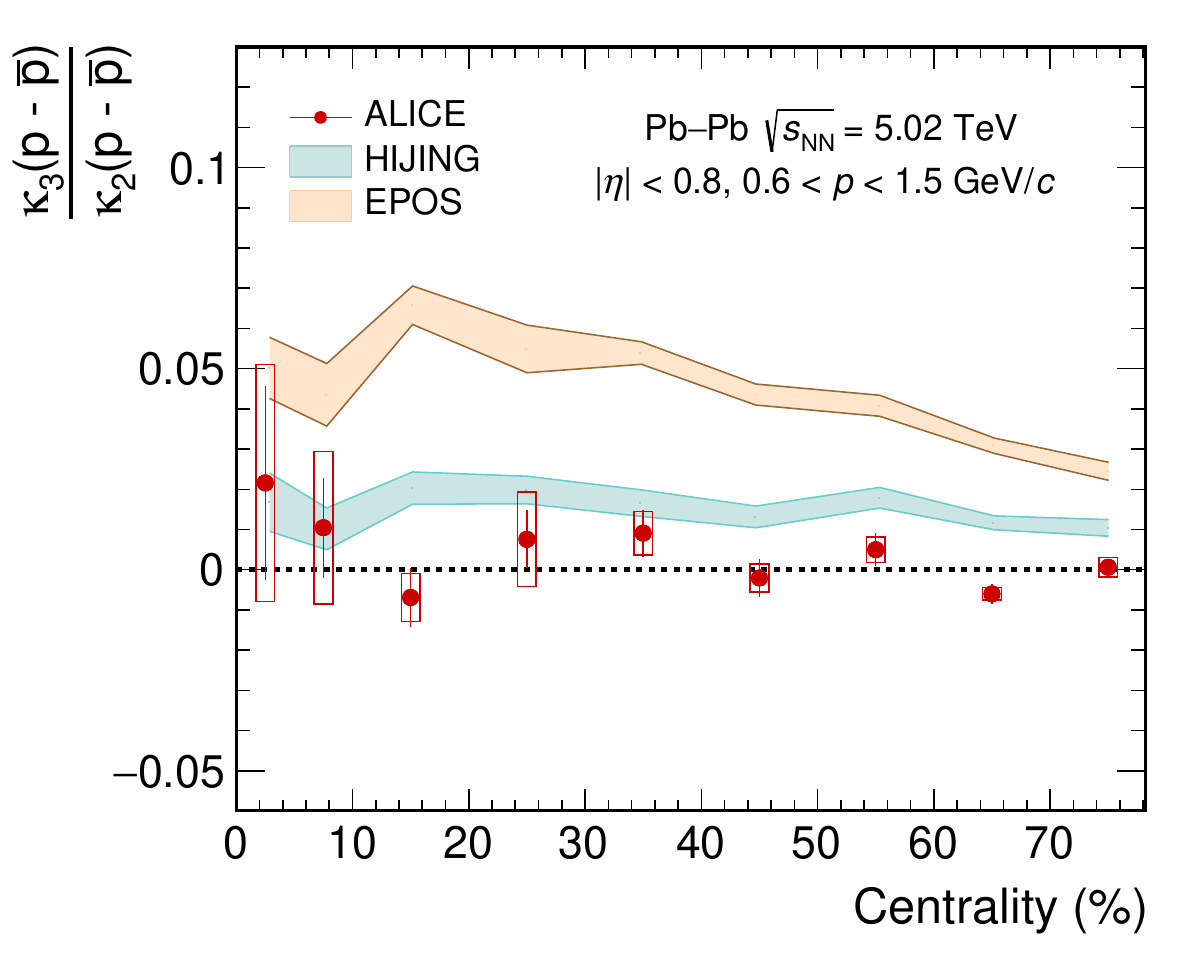}
     \end{minipage}%
     \begin{minipage}{.5\textwidth}
         \vspace*{\fill}
         \centering
         \includegraphics[width=\textwidth]{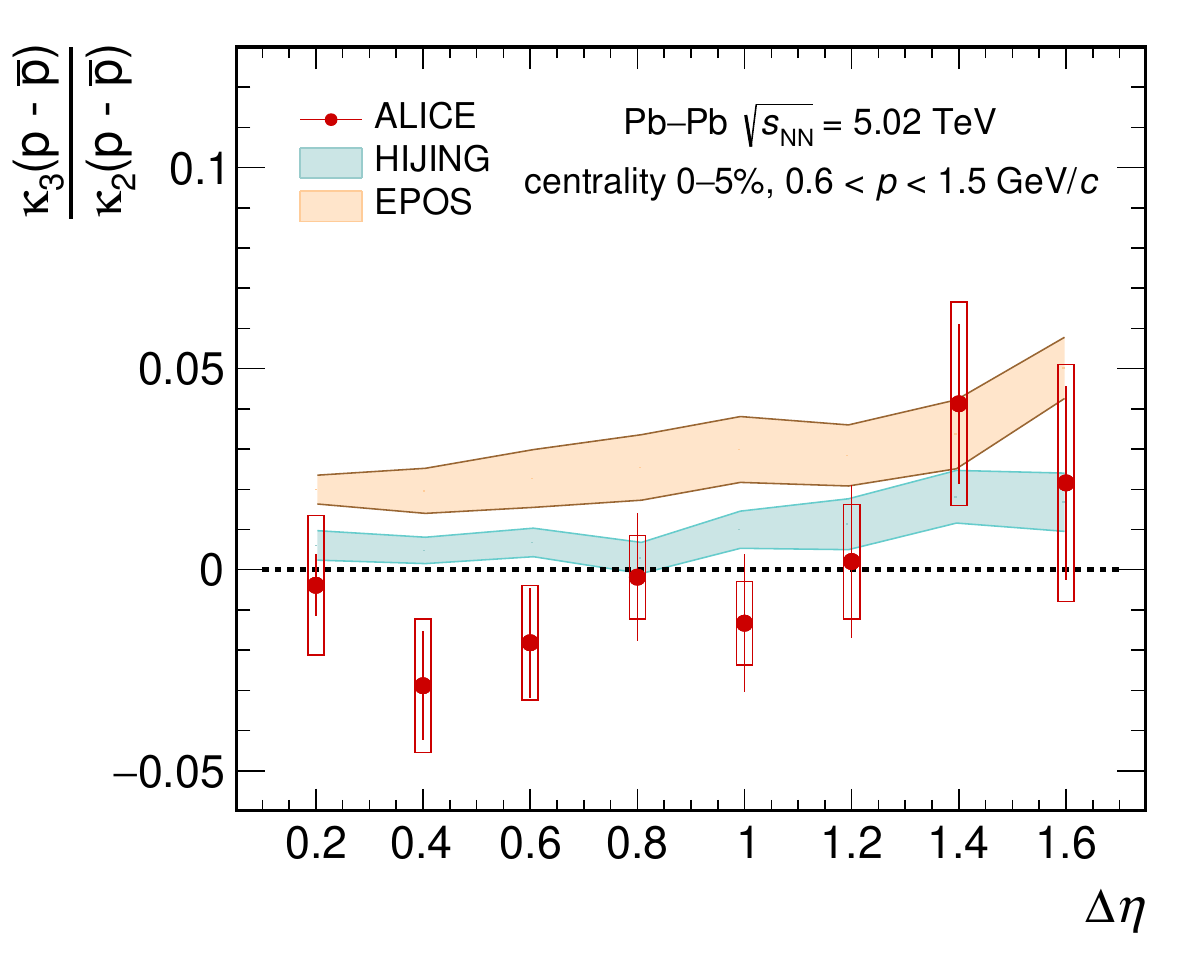}
     \end{minipage}%
  \caption{(Color online) Centrality (left) and pseudorapidity interval (right) dependence of the ratio of third- to second-order cumulants for net protons at $\snn=5.02$~TeV. The ALICE data are shown by red markers, while the colored shaded bands represent the results from HIJING~\cite{Gyulassy:1994ew} and EPOS~\cite{Pierog:2013ria} model calculations.}
  \label{fig:3rd_CentDep}
\end{figure}

In \figref{fig:3rd_CentDep}, the third-order cumulant measurements are also compared with HIJING and EPOS model calculation results. Both models include baryon number conservation but, as mentioned above, the net-proton number is positive within the current experimental acceptance. Therefore, the resulting third-order cumulants for all centrality and pseudorapidity difference intervals shift toward positive values and are affected by the volume fluctuations~\cite{Braun-Munzinger:2016yjz} visible in the 10--20\% centrality interval, where the centrality range doubles (left panel). The consistency of the experimental results of the third- to second-order cumulant ratio with a value of zero, which is lower than the expectation from the EPOS and HIJING generators, is a confirmation that the proton-to-antiproton ratio is closer to unity than calculated by these generators and that the systematic uncertainties for these measurements are under good control. A value of $\kappa_{3}/\kappa_{2}$ consistent with zero within small uncertainties also indicates that $\mu_{\rm B}$ is very close to zero at LHC energies.

\section{Conclusions} \label{Summary}

In summary, net-proton cumulant measurements up to third order and net-pion and net-kaon second-order cumulant measurements are reported. The technical challenges related to data analysis, in particular efficiency correction and event pile-up, are overcome as discussed in detail. Resonance contributions prove to be challenging in the study of fluctuations of the net-electric charge and the net-strangeness. A deviation of about 4\% from the Skellam baseline is observed for the second-order net-proton cumulants for the widest $\Delta\eta$ interval. Investigation of this deviation in light of baryon number conservation led to the conclusion that the 2010 data from ALICE~\cite{ALICE:2019nbs} indicate the presence of long-range rapidity correlations between protons and antiprotons originating from the early phase of the collision. This finding is corroborated by the present analysis including the higher luminosity 2015 data with significantly different experimental conditions. Results of calculations using the HIJING generator, based on the Lund string model, reflect a much smaller correlation length of one unit of rapidity. This observed discrepancy calls into question the mechanism implemented in the Lund string model for the production of baryons. Baryon production as implemented in the EPOS event generator reflects the long-range correlation observed in the data. After accounting for the effect of baryon number conservation, the data from ALICE are consistent with LQCD expectations up to the third-order cumulants of the net protons. The finding of third-order net-proton cumulants consistent with zero with a precision of better than 4\% is promising for the analysis of the higher-order cumulants during the operation of LHC with increased Pb--Pb luminosity~\cite{Citron:2018lsq} starting in 2022 and for the future heavy-ion detector planned for the early 2030s~\cite{ALICE:2803563}.

\newenvironment{acknowledgement}{\relax}{\relax}
\begin{acknowledgement}
\section*{Acknowledgements}

The ALICE Collaboration would like to thank all its engineers and technicians for their invaluable contributions to the construction of the experiment and the CERN accelerator teams for the outstanding performance of the LHC complex.
The ALICE Collaboration gratefully acknowledges the resources and support provided by all Grid centres and the Worldwide LHC Computing Grid (WLCG) collaboration.
The ALICE Collaboration acknowledges the following funding agencies for their support in building and running the ALICE detector:
A. I. Alikhanyan National Science Laboratory (Yerevan Physics Institute) Foundation (ANSL), State Committee of Science and World Federation of Scientists (WFS), Armenia;
Austrian Academy of Sciences, Austrian Science Fund (FWF): [M 2467-N36] and Nationalstiftung f\"{u}r Forschung, Technologie und Entwicklung, Austria;
Ministry of Communications and High Technologies, National Nuclear Research Center, Azerbaijan;
Conselho Nacional de Desenvolvimento Cient\'{\i}fico e Tecnol\'{o}gico (CNPq), Financiadora de Estudos e Projetos (Finep), Funda\c{c}\~{a}o de Amparo \`{a} Pesquisa do Estado de S\~{a}o Paulo (FAPESP) and Universidade Federal do Rio Grande do Sul (UFRGS), Brazil;
Bulgarian Ministry of Education and Science, within the National Roadmap for Research Infrastructures 2020-2027 (object CERN), Bulgaria;
Ministry of Education of China (MOEC) , Ministry of Science \& Technology of China (MSTC) and National Natural Science Foundation of China (NSFC), China;
Ministry of Science and Education and Croatian Science Foundation, Croatia;
Centro de Aplicaciones Tecnol\'{o}gicas y Desarrollo Nuclear (CEADEN), Cubaenerg\'{\i}a, Cuba;
Ministry of Education, Youth and Sports of the Czech Republic, Czech Republic;
The Danish Council for Independent Research | Natural Sciences, the VILLUM FONDEN and Danish National Research Foundation (DNRF), Denmark;
Helsinki Institute of Physics (HIP), Finland;
Commissariat \`{a} l'Energie Atomique (CEA) and Institut National de Physique Nucl\'{e}aire et de Physique des Particules (IN2P3) and Centre National de la Recherche Scientifique (CNRS), France;
Bundesministerium f\"{u}r Bildung und Forschung (BMBF) and GSI Helmholtzzentrum f\"{u}r Schwerionenforschung GmbH, Germany;
General Secretariat for Research and Technology, Ministry of Education, Research and Religions, Greece;
National Research, Development and Innovation Office, Hungary;
Department of Atomic Energy Government of India (DAE), Department of Science and Technology, Government of India (DST), University Grants Commission, Government of India (UGC) and Council of Scientific and Industrial Research (CSIR), India;
National Research and Innovation Agency - BRIN, Indonesia;
Istituto Nazionale di Fisica Nucleare (INFN), Italy;
Japanese Ministry of Education, Culture, Sports, Science and Technology (MEXT) and Japan Society for the Promotion of Science (JSPS) KAKENHI, Japan;
Consejo Nacional de Ciencia (CONACYT) y Tecnolog\'{i}a, through Fondo de Cooperaci\'{o}n Internacional en Ciencia y Tecnolog\'{i}a (FONCICYT) and Direcci\'{o}n General de Asuntos del Personal Academico (DGAPA), Mexico;
Nederlandse Organisatie voor Wetenschappelijk Onderzoek (NWO), Netherlands;
The Research Council of Norway, Norway;
Commission on Science and Technology for Sustainable Development in the South (COMSATS), Pakistan;
Pontificia Universidad Cat\'{o}lica del Per\'{u}, Peru;
Ministry of Education and Science, National Science Centre and WUT ID-UB, Poland;
Korea Institute of Science and Technology Information and National Research Foundation of Korea (NRF), Republic of Korea;
Ministry of Education and Scientific Research, Institute of Atomic Physics, Ministry of Research and Innovation and Institute of Atomic Physics and University Politehnica of Bucharest, Romania;
Ministry of Education, Science, Research and Sport of the Slovak Republic, Slovakia;
National Research Foundation of South Africa, South Africa;
Swedish Research Council (VR) and Knut \& Alice Wallenberg Foundation (KAW), Sweden;
European Organization for Nuclear Research, Switzerland;
Suranaree University of Technology (SUT), National Science and Technology Development Agency (NSTDA), Thailand Science Research and Innovation (TSRI) and National Science, Research and Innovation Fund (NSRF), Thailand;
Turkish Energy, Nuclear and Mineral Research Agency (TENMAK), Turkey;
National Academy of  Sciences of Ukraine, Ukraine;
Science and Technology Facilities Council (STFC), United Kingdom;
National Science Foundation of the United States of America (NSF) and United States Department of Energy, Office of Nuclear Physics (DOE NP), United States of America.
In addition, individual groups or members have received support from:
Marie Sk\l{}odowska Curie, European Research Council, Strong 2020 - Horizon 2020 (grant nos. 950692, 824093, 896850), European Union;
Academy of Finland (Center of Excellence in Quark Matter) (grant nos. 346327, 346328), Finland;
Programa de Apoyos para la Superaci\'{o}n del Personal Acad\'{e}mico, UNAM, Mexico.

\end{acknowledgement}

\bibliographystyle{utphys}   
\bibliography{bibliography}

\providecommand{\href}[2]{#2}\begingroup\raggedright\begin{thebibliography}{10}

\bibitem{Shuryak:1980tp}
E.~V. Shuryak, ``{Quantum Chromodynamics and the Theory of Superdense
  Matter}'', \href{http://dx.doi.org/10.1016/0370-1573(80)90105-2}{{\em Phys.
  Rept.} {\bfseries 61} (1980) 71}.

\bibitem{Ding:2015ona}
H.-T. Ding, F.~Karsch, and S.~Mukherjee, ``{Thermodynamics of
  strong-interaction matter from Lattice QCD}'',
  \href{http://dx.doi.org/10.1142/S0218301315300076}{{\em Int. J. Mod. Phys. E}
  {\bfseries 24} (2015) 1530007},
  \href{http://arxiv.org/abs/1504.05274}{{\ttfamily arXiv:1504.05274
  [hep-lat]}}.

\bibitem{Borsanyi:2013bia}
S.~Borsanyi, Z.~Fodor, C.~Hoelbling, S.~D. Katz, S.~Krieg, and K.~K. Szabo,
  ``{Full result for the QCD equation of state with 2+1 flavors}'',
  \href{http://dx.doi.org/10.1016/j.physletb.2014.01.007}{{\em Phys. Lett. B}
  {\bfseries 730} (2014) 99}, \href{http://arxiv.org/abs/1309.5258}{{\ttfamily
  arXiv:1309.5258 [hep-lat]}}.

\bibitem{Andronic:2017pug}
A.~Andronic, P.~Braun-Munzinger, K.~Redlich, and J.~Stachel, ``{Decoding the
  phase structure of QCD via particle production at high energy}'',
  \href{http://dx.doi.org/10.1038/s41586-018-0491-6}{{\em Nature} {\bfseries
  561} (2018) 321}, \href{http://arxiv.org/abs/1710.09425}{{\ttfamily
  arXiv:1710.09425 [nucl-th]}}.

\bibitem{HotQCD:2018pds}
{\bfseries HotQCD} Collaboration, A.~Bazavov {\em et~al.}, ``{Chiral crossover
  in QCD at zero and non-zero chemical potentials}'',
  \href{http://dx.doi.org/10.1016/j.physletb.2019.05.013}{{\em Phys. Lett. B}
  {\bfseries 795} (2019) 15}, \href{http://arxiv.org/abs/1812.08235}{{\ttfamily
  arXiv:1812.08235 [hep-lat]}}.

\bibitem{Borsanyi:2020fev}
S.~Borsanyi, Z.~Fodor, J.~N. Guenther, R.~Kara, S.~D. Katz, P.~Parotto,
  A.~Pasztor, C.~Ratti, and K.~K. Szabo, ``{QCD Crossover at Finite Chemical
  Potential from Lattice Simulations}'',
  \href{http://dx.doi.org/10.1103/PhysRevLett.125.052001}{{\em Phys. Rev.
  Lett.} {\bfseries 125} (2020) 052001},
  \href{http://arxiv.org/abs/2002.02821}{{\ttfamily arXiv:2002.02821
  [hep-lat]}}.

\bibitem{Pisarski:1983ms}
R.~D. Pisarski and F.~Wilczek, ``{Remarks on the Chiral Phase Transition in
  Chromodynamics}'', \href{http://dx.doi.org/10.1103/PhysRevD.29.338}{{\em
  Phys. Rev. D} {\bfseries 29} (1984) 338}.

\bibitem{Luo:2017faz}
X.~Luo and N.~Xu, ``{Search for the QCD Critical Point with Fluctuations of
  Conserved Quantities in Relativistic Heavy-Ion Collisions at RHIC : An
  Overview}'', \href{http://dx.doi.org/10.1007/s41365-017-0257-0}{{\em Nucl.
  Sci. Tech.} {\bfseries 28} (2017) 112},
  \href{http://arxiv.org/abs/1701.02105}{{\ttfamily arXiv:1701.02105
  [nucl-ex]}}.

\bibitem{Xu:2018vnf}
N.~Xu, ``{Exploration of the QCD Phase Diagram at Finite Baryon Density Region:
  Recent Results from RHIC Beam Energy Scan-I}'',
  \href{http://dx.doi.org/10.1007/978-3-319-73171-1_1}{{\em Springer Proc.
  Phys.} {\bfseries 203} (2018) 1}.

\bibitem{NA49:2009diu}
{\bfseries NA49} Collaboration, T.~Anticic {\em et~al.}, ``{Search for the QCD
  critical point in nuclear collisions at the CERN SPS}'',
  \href{http://dx.doi.org/10.1103/PhysRevC.81.064907}{{\em Phys. Rev. C}
  {\bfseries 81} (2010) 064907},
  \href{http://arxiv.org/abs/0912.4198}{{\ttfamily arXiv:0912.4198 [nucl-ex]}}.

\bibitem{CBM:2016kpk}
{\bfseries CBM} Collaboration, T.~Ablyazimov {\em et~al.}, ``{Challenges in QCD
  matter physics --The scientific programme of the Compressed Baryonic Matter
  experiment at FAIR}'',
  \href{http://dx.doi.org/10.1140/epja/i2017-12248-y}{{\em Eur. Phys. J. A}
  {\bfseries 53} (2017) 60}, \href{http://arxiv.org/abs/1607.01487}{{\ttfamily
  arXiv:1607.01487 [nucl-ex]}}.

\bibitem{HotQCD:2019xnw}
{\bfseries HotQCD} Collaboration, H.~T. Ding {\em et~al.}, ``{Chiral Phase
  Transition Temperature in ( 2+1 )-Flavor QCD}'',
  \href{http://dx.doi.org/10.1103/PhysRevLett.123.062002}{{\em Phys. Rev.
  Lett.} {\bfseries 123} (2019) 062002},
  \href{http://arxiv.org/abs/1903.04801}{{\ttfamily arXiv:1903.04801
  [hep-lat]}}.

\bibitem{Borsanyi:2018grb}
S.~Borsanyi, Z.~Fodor, J.~N. Guenther, S.~K. Katz, K.~K. Szabo, A.~Pasztor,
  I.~Portillo, and C.~Ratti, ``{Higher order fluctuations and correlations of
  conserved charges from lattice QCD}'',
  \href{http://dx.doi.org/10.1007/JHEP10(2018)205}{{\em JHEP} {\bfseries 10}
  (2018) 205}, \href{http://arxiv.org/abs/1805.04445}{{\ttfamily
  arXiv:1805.04445 [hep-lat]}}.

\bibitem{Bazavov:2011nk}
A.~Bazavov {\em et~al.}, ``{The chiral and deconfinement aspects of the QCD
  transition}'', \href{http://dx.doi.org/10.1103/PhysRevD.85.054503}{{\em Phys.
  Rev. D} {\bfseries 85} (2012) 054503},
  \href{http://arxiv.org/abs/1111.1710}{{\ttfamily arXiv:1111.1710 [hep-lat]}}.

\bibitem{ALICE:2013jfw}
{\bfseries ALICE} Collaboration, E.~Abbas {\em et~al.}, ``{Centrality
  dependence of the pseudorapidity density distribution for charged particles
  in Pb-Pb collisions at $\sqrt{s_{\rm NN}}$ = 2.76 TeV}'',
  \href{http://dx.doi.org/10.1016/j.physletb.2013.09.022}{{\em Phys. Lett. B}
  {\bfseries 726} (2013) 610}, \href{http://arxiv.org/abs/1304.0347}{{\ttfamily
  arXiv:1304.0347 [nucl-ex]}}.

\bibitem{Friman:2011pf}
B.~Friman, F.~Karsch, K.~Redlich, and V.~Skokov, ``{Fluctuations as probe of
  the {QCD} phase transition and freeze-out in heavy ion collisions at {LHC}
  and {RHIC}}'', \href{http://dx.doi.org/10.1140/epjc/s10052-011-1694-2}{{\em
  Eur. Phys. J. C} {\bfseries 71} (2011) 1694},
  \href{http://arxiv.org/abs/1103.3511}{{\ttfamily arXiv:1103.3511 [hep-ph]}}.

\bibitem{ParisenToldin:2003hq}
F.~Parisen~Toldin, A.~Pelissetto, and E.~Vicari, ``{The 3-D O(4) universality
  class and the phase transition in two flavor QCD}'',
  \href{http://dx.doi.org/10.1088/1126-6708/2003/07/029}{{\em JHEP} {\bfseries
  07} (2003) 029}, \href{http://arxiv.org/abs/hep-ph/0305264}{{\ttfamily
  arXiv:hep-ph/0305264}}.

\bibitem{Braun-Munzinger:2016yjz}
P.~Braun-Munzinger, A.~Rustamov, and J.~Stachel, ``{Bridging the gap between
  event-by-event fluctuation measurements and theory predictions in
  relativistic nuclear collisions}'',
  \href{http://dx.doi.org/10.1016/j.nuclphysa.2017.01.011}{{\em Nucl. Phys. A}
  {\bfseries 960} (2017) 114},
  \href{http://arxiv.org/abs/1612.00702}{{\ttfamily arXiv:1612.00702
  [nucl-th]}}.

\bibitem{Gavai:2010zn}
R.~V. Gavai and S.~Gupta, ``{Lattice QCD predictions for shapes of event
  distributions along the freezeout curve in heavy-ion collisions}'',
  \href{http://dx.doi.org/10.1016/j.physletb.2011.01.006}{{\em Phys. Lett. B}
  {\bfseries 696} (2011) 459}, \href{http://arxiv.org/abs/1001.3796}{{\ttfamily
  arXiv:1001.3796 [hep-lat]}}.

\bibitem{Ejiri:2005wq}
S.~Ejiri, F.~Karsch, and K.~Redlich, ``{Hadronic fluctuations at the QCD phase
  transition}'', \href{http://dx.doi.org/10.1016/j.physletb.2005.11.083}{{\em
  Phys. Lett. B} {\bfseries 633} (2006) 275},
  \href{http://arxiv.org/abs/hep-ph/0509051}{{\ttfamily arXiv:hep-ph/0509051}}.

\bibitem{Bzdak:2012an}
A.~Bzdak, V.~Koch, and V.~Skokov, ``{Baryon number conservation and the
  cumulants of the net proton distribution}'',
  \href{http://dx.doi.org/10.1103/PhysRevC.87.014901}{{\em Phys. Rev. C}
  {\bfseries 87} (2013) 014901},
  \href{http://arxiv.org/abs/1203.4529}{{\ttfamily arXiv:1203.4529 [hep-ph]}}.

\bibitem{Braun-Munzinger:2019yxj}
P.~Braun-Munzinger, A.~Rustamov, and J.~Stachel, ``{The role of the local
  conservation laws in fluctuations of conserved charges}'',
  \href{http://arxiv.org/abs/1907.03032}{{\ttfamily arXiv:1907.03032
  [nucl-th]}}.

\bibitem{Vovchenko:2020kwg}
V.~Vovchenko and V.~Koch, ``{Particlization of an interacting hadron resonance
  gas with global conservation laws for event-by-event fluctuations in
  heavy-ion collisions}'',
  \href{http://dx.doi.org/10.1103/PhysRevC.103.044903}{{\em Phys. Rev. C}
  {\bfseries 103} (2021) 044903},
  \href{http://arxiv.org/abs/2012.09954}{{\ttfamily arXiv:2012.09954
  [hep-ph]}}.

\bibitem{Ohnishi:2016bdf}
Y.~Ohnishi, M.~Kitazawa, and M.~Asakawa, ``{Thermal blurring of event-by-event
  fluctuations generated by rapidity conversion}'',
  \href{http://dx.doi.org/10.1103/PhysRevC.94.044905}{{\em Phys. Rev. C}
  {\bfseries 94} (2016) 044905},
  \href{http://arxiv.org/abs/1606.03827}{{\ttfamily arXiv:1606.03827
  [nucl-th]}}.

\bibitem{Bluhm:2020mpc}
M.~Bluhm {\em et~al.}, ``{Dynamics of critical fluctuations: Theory
  \textendash{} phenomenology \textendash{} heavy-ion collisions}'',
  \href{http://dx.doi.org/10.1016/j.nuclphysa.2020.122016}{{\em Nucl. Phys. A}
  {\bfseries 1003} (2020) 122016},
  \href{http://arxiv.org/abs/2001.08831}{{\ttfamily arXiv:2001.08831
  [nucl-th]}}.

\bibitem{Shen:2017bsr}
C.~Shen and B.~Schenke, ``{Dynamical initial state model for relativistic
  heavy-ion collisions}'',
  \href{http://dx.doi.org/10.1103/PhysRevC.97.024907}{{\em Phys. Rev. C}
  {\bfseries 97} (2018) 024907},
  \href{http://arxiv.org/abs/1710.00881}{{\ttfamily arXiv:1710.00881
  [nucl-th]}}.

\bibitem{Savchuk:2021aog}
O.~Savchuk, V.~Vovchenko, V.~Koch, J.~Steinheimer, and H.~Stoecker,
  ``{Constraining baryon annihilation in the hadronic phase of heavy-ion
  collisions via event-by-event fluctuations}'',
  \href{http://dx.doi.org/10.1016/j.physletb.2022.136983}{{\em Phys. Lett. B}
  {\bfseries 827} (2022) 136983},
  \href{http://arxiv.org/abs/2106.08239}{{\ttfamily arXiv:2106.08239
  [hep-ph]}}.

\bibitem{ALICE:2008ngc}
{\bfseries ALICE} Collaboration, K.~Aamodt {\em et~al.}, ``{The ALICE
  experiment at the CERN LHC}'',
  \href{http://dx.doi.org/10.1088/1748-0221/3/08/S08002}{{\em JINST} {\bfseries
  3} (2008) S08002}.

\bibitem{ALICE:2014sbx}
{\bfseries ALICE} Collaboration, B.~B. Abelev {\em et~al.}, ``{Performance of
  the ALICE Experiment at the CERN LHC}'',
  \href{http://dx.doi.org/10.1142/S0217751X14300440}{{\em Int. J. Mod. Phys. A}
  {\bfseries 29} (2014) 1430044},
  \href{http://arxiv.org/abs/1402.4476}{{\ttfamily arXiv:1402.4476 [nucl-ex]}}.

\bibitem{ALICE:2019nbs}
{\bfseries ALICE} Collaboration, S.~Acharya {\em et~al.}, ``{Global baryon
  number conservation encoded in net-proton fluctuations measured in Pb-Pb
  collisions at $\sqrt{s_{\rm NN}}$ = 2.76 TeV}'',
  \href{http://dx.doi.org/10.1016/j.physletb.2020.135564}{{\em Phys. Lett. B}
  {\bfseries 807} (2020) 135564},
  \href{http://arxiv.org/abs/1910.14396}{{\ttfamily arXiv:1910.14396
  [nucl-ex]}}.

\bibitem{ALICE:2017jsh}
{\bfseries ALICE} Collaboration, S.~Acharya {\em et~al.}, ``{Relative particle
  yield fluctuations in $\text{ Pb-Pb }$ collisions at $\sqrt{s_\mathrm{{NN}}}
  =2.76\hbox { TeV}$}'',
  \href{http://dx.doi.org/10.1140/epjc/s10052-019-6711-x}{{\em Eur. Phys. J. C}
  {\bfseries 79} (2019) 236}, \href{http://arxiv.org/abs/1712.07929}{{\ttfamily
  arXiv:1712.07929 [nucl-ex]}}.

\bibitem{ALICE:2013axi}
{\bfseries ALICE} Collaboration, E.~Abbas {\em et~al.}, ``{Performance of the
  ALICE VZERO system}'',
  \href{http://dx.doi.org/10.1088/1748-0221/8/10/P10016}{{\em JINST} {\bfseries
  8} (2013) P10016}, \href{http://arxiv.org/abs/1306.3130}{{\ttfamily
  arXiv:1306.3130 [nucl-ex]}}.

\bibitem{ALICE:2013hur}
{\bfseries ALICE} Collaboration, B.~Abelev {\em et~al.}, ``{Centrality
  determination of Pb-Pb collisions at $\sqrt{s_{\rm NN}}$ = 2.76 TeV with
  ALICE}'', \href{http://dx.doi.org/10.1103/PhysRevC.88.044909}{{\em Phys. Rev.
  C} {\bfseries 88} (2013) 044909},
  \href{http://arxiv.org/abs/1301.4361}{{\ttfamily arXiv:1301.4361 [nucl-ex]}}.

\bibitem{Loizides:2014vua}
C.~Loizides, J.~Nagle, and P.~Steinberg, ``{Improved version of the PHOBOS
  Glauber Monte Carlo}'',
  \href{http://dx.doi.org/10.1016/j.softx.2015.05.001}{{\em SoftwareX}
  {\bfseries 1-2} (2015) 13}, \href{http://arxiv.org/abs/1408.2549}{{\ttfamily
  arXiv:1408.2549 [nucl-ex]}}.

\bibitem{Miller:2007ri}
M.~L. Miller, K.~Reygers, S.~J. Sanders, and P.~Steinberg, ``{Glauber modeling
  in high energy nuclear collisions}'',
  \href{http://dx.doi.org/10.1146/annurev.nucl.57.090506.123020}{{\em Ann. Rev.
  Nucl. Part. Sci.} {\bfseries 57} (2007) 205},
  \href{http://arxiv.org/abs/nucl-ex/0701025}{{\ttfamily
  arXiv:nucl-ex/0701025}}.

\bibitem{Alme:2010ke}
J.~Alme {\em et~al.}, ``{The ALICE TPC, a large 3-dimensional tracking device
  with fast readout for ultra-high multiplicity events}'',
  \href{http://dx.doi.org/10.1016/j.nima.2010.04.042}{{\em Nucl. Instrum. Meth.
  A} {\bfseries 622} (2010) 316},
  \href{http://arxiv.org/abs/1001.1950}{{\ttfamily arXiv:1001.1950
  [physics.ins-det]}}.

\bibitem{CERN-LHCC-99-012}
{\bfseries ALICE} Collaboration, ``{ALICE Inner Tracking System (ITS):
  Technical Design Report}'', {\em CERN-LHCC-99-012} (1999) .
  \url{http://cds.cern.ch/record/391175}.

\bibitem{ALICE:2010tia}
{\bfseries ALICE} Collaboration, K.~Aamodt {\em et~al.}, ``{Alignment of the
  ALICE Inner Tracking System with cosmic-ray tracks}'',
  \href{http://dx.doi.org/10.1088/1748-0221/5/03/P03003}{{\em JINST} {\bfseries
  5} (2010) P03003}, \href{http://arxiv.org/abs/1001.0502}{{\ttfamily
  arXiv:1001.0502 [physics.ins-det]}}.

\bibitem{ALICE:1999edx}
{\bfseries ALICE} Collaboration, G.~Dellacasa {\em et~al.}, ``{ALICE technical
  design report of the zero degree calorimeter (ZDC)}'', {\em CERN-LHCC-99-005}
  (1999) . \url{https://cds.cern.ch/record/381433}.

\bibitem{Arslandok:2022dyb}
M.~Arslandok, E.~Hellb\"ar, M.~Ivanov, R.~H. M\"unzer, and J.~Wiechula,
  ``{Track Reconstruction in a High-Density Environment with ALICE}'',
  \href{http://dx.doi.org/10.3390/particles5010008}{{\em Particles} {\bfseries
  5} (2022) 84}, \href{http://arxiv.org/abs/2203.10325}{{\ttfamily
  arXiv:2203.10325 [physics.ins-det]}}.

\bibitem{ALICE:2010yje}
{\bfseries ALICE} Collaboration, K.~Aamodt {\em et~al.}, ``{Suppression of
  Charged Particle Production at Large Transverse Momentum in Central Pb-Pb
  Collisions at $\sqrt{s_{\rm NN}} =$ 2.76 TeV}'',
  \href{http://dx.doi.org/10.1016/j.physletb.2010.12.020}{{\em Phys. Lett. B}
  {\bfseries 696} (2011) 30}, \href{http://arxiv.org/abs/1012.1004}{{\ttfamily
  arXiv:1012.1004 [nucl-ex]}}.

\bibitem{Gazdzicki:2011xz}
M.~Gazdzicki, K.~Grebieszkow, M.~Mackowiak, and S.~Mrowczynski, ``{Identity
  method to study chemical fluctuations in relativistic heavy-ion
  collisions}'', \href{http://dx.doi.org/10.1103/PhysRevC.83.054907}{{\em Phys.
  Rev. C} {\bfseries 83} (2011) 054907},
  \href{http://arxiv.org/abs/1103.2887}{{\ttfamily arXiv:1103.2887 [nucl-th]}}.

\bibitem{Rustamov:2012bx}
A.~Rustamov and M.~I. Gorenstein, ``{Identity method for the determination of
  the moments of multiplicity distributions}'',
  \href{http://dx.doi.org/10.1103/PhysRevC.86.044906}{{\em Phys. Rev. C}
  {\bfseries 86} (2012) 044906},
  \href{http://arxiv.org/abs/1204.6632}{{\ttfamily arXiv:1204.6632 [nucl-th]}}.

\bibitem{Arslandok:2018pcu}
M.~Arslandok and A.~Rustamov, ``{TIdentity module for the reconstruction of the
  moments of multiplicity distributions}'',
  \href{http://dx.doi.org/10.1016/j.nima.2019.162622}{{\em Nucl. Instrum. Meth.
  A} {\bfseries 946} (2019) 162622},
  \href{http://arxiv.org/abs/1807.06370}{{\ttfamily arXiv:1807.06370
  [hep-ex]}}.

\bibitem{Bzdak:2012ab}
A.~Bzdak and V.~Koch, ``{Acceptance corrections to net baryon and net charge
  cumulants}'', \href{http://dx.doi.org/10.1103/PhysRevC.86.044904}{{\em Phys.
  Rev. C} {\bfseries 86} (2012) 044904},
  \href{http://arxiv.org/abs/1206.4286}{{\ttfamily arXiv:1206.4286 [nucl-th]}}.

\bibitem{Bzdak:2013pha}
A.~Bzdak and V.~Koch, ``{Local Efficiency Corrections to Higher Order
  Cumulants}'', \href{http://dx.doi.org/10.1103/PhysRevC.91.027901}{{\em Phys.
  Rev. C} {\bfseries 91} (2015) 027901},
  \href{http://arxiv.org/abs/1312.4574}{{\ttfamily arXiv:1312.4574 [nucl-th]}}.

\bibitem{Nonaka:2017kko}
T.~Nonaka, M.~Kitazawa, and S.~Esumi, ``{More efficient formulas for efficiency
  correction of cumulants and effect of using averaged efficiency}'',
  \href{http://dx.doi.org/10.1103/PhysRevC.95.064912}{{\em Phys. Rev. C}
  {\bfseries 95} (2017) 064912},
  \href{http://arxiv.org/abs/1702.07106}{{\ttfamily arXiv:1702.07106
  [physics.data-an]}}. [Erratum: Phys.Rev.C 103, 029901 (2021)].

\bibitem{Pandav:2018bdx}
A.~Pandav, D.~Mallick, and B.~Mohanty, ``{Effect of limited statistics on
  higher order cumulants measurement in heavy-ion collision experiments}'',
  \href{http://dx.doi.org/10.1016/j.nuclphysa.2019.08.002}{{\em Nucl. Phys. A}
  {\bfseries 991} (2019) 121608},
  \href{http://arxiv.org/abs/1809.08892}{{\ttfamily arXiv:1809.08892
  [nucl-ex]}}.

\bibitem{GEANT4:2002zbu}
{\bfseries GEANT4} Collaboration, S.~Agostinelli {\em et~al.}, ``{GEANT4--a
  simulation toolkit}'',
  \href{http://dx.doi.org/10.1016/S0168-9002(03)01368-8}{{\em Nucl. Instrum.
  Meth. A} {\bfseries 506} (2003) 250--303}.

\bibitem{Gyulassy:1994ew}
M.~Gyulassy and X.-N. Wang, ``{HIJING 1.0: A Monte Carlo program for parton and
  particle production in high-energy hadronic and nuclear collisions}'',
  \href{http://dx.doi.org/10.1016/0010-4655(94)90057-4}{{\em Comput. Phys.
  Commun.} {\bfseries 83} (1994) 307},
  \href{http://arxiv.org/abs/nucl-th/9502021}{{\ttfamily
  arXiv:nucl-th/9502021}}.

\bibitem{Kitazawa:2012at}
M.~Kitazawa and M.~Asakawa, ``{Relation between baryon number fluctuations and
  experimentally observed proton number fluctuations in relativistic heavy ion
  collisions}'', \href{http://dx.doi.org/10.1103/PhysRevC.86.024904}{{\em Phys.
  Rev. C} {\bfseries 86} (2012) 024904},
  \href{http://arxiv.org/abs/1205.3292}{{\ttfamily arXiv:1205.3292 [nucl-th]}}.
  [Erratum: Phys.Rev.C 86, 069902 (2012)].

\bibitem{Braun-Munzinger:2011xux}
P.~Braun-Munzinger, B.~Friman, F.~Karsch, K.~Redlich, and V.~Skokov,
  ``{Net-charge probability distributions in heavy ion collisions at chemical
  freeze-out}'', \href{http://dx.doi.org/10.1016/j.nuclphysa.2012.02.010}{{\em
  Nucl. Phys. A} {\bfseries 880} (2012) 48},
  \href{http://arxiv.org/abs/1111.5063}{{\ttfamily arXiv:1111.5063 [hep-ph]}}.

\bibitem{Skellam}
J.~G. Skellam, ``{The frequency distribution of the difference between two
  Poisson variates belonging to different populations}'', {\em {J. Royal Stat.
  Soc.}} {\bfseries A109(3)} (1946) 296.

\bibitem{ALICE:2013mez}
{\bfseries ALICE} Collaboration, B.~Abelev {\em et~al.}, ``{Centrality
  dependence of $\pi$, K, p production in Pb-Pb collisions at $\sqrt{s_{\rm
  NN}}$ = 2.76 TeV}'', \href{http://dx.doi.org/10.1103/PhysRevC.88.044910}{{\em
  Phys. Rev. C} {\bfseries 88} (2013) 044910},
  \href{http://arxiv.org/abs/1303.0737}{{\ttfamily arXiv:1303.0737 [hep-ex]}}.

\bibitem{Bazavov:2020bjn}
A.~Bazavov {\em et~al.}, ``{Skewness, kurtosis, and the fifth and sixth order
  cumulants of net baryon-number distributions from lattice QCD confront
  high-statistics STAR data}'',
  \href{http://dx.doi.org/10.1103/PhysRevD.101.074502}{{\em Phys. Rev. D}
  {\bfseries 101} (2020) 074502},
  \href{http://arxiv.org/abs/2001.08530}{{\ttfamily arXiv:2001.08530
  [hep-lat]}}.

\bibitem{Braun-Munzinger:2020jbk}
P.~Braun-Munzinger, B.~Friman, K.~Redlich, A.~Rustamov, and J.~Stachel,
  ``{Relativistic nuclear collisions: Establishing a non-critical baseline for
  fluctuation measurements}'',
  \href{http://dx.doi.org/10.1016/j.nuclphysa.2021.122141}{{\em Nucl. Phys. A}
  {\bfseries 1008} (2021) 122141},
  \href{http://arxiv.org/abs/2007.02463}{{\ttfamily arXiv:2007.02463
  [nucl-th]}}.

\bibitem{Pierog:2013ria}
T.~Pierog, I.~Karpenko, J.~M. Katzy, E.~Yatsenko, and K.~Werner, ``{EPOS LHC:
  Test of collective hadronization with data measured at the CERN Large Hadron
  Collider}'', \href{http://dx.doi.org/10.1103/PhysRevC.92.034906}{{\em Phys.
  Rev. C} {\bfseries 92} (2015) 034906},
  \href{http://arxiv.org/abs/1306.0121}{{\ttfamily arXiv:1306.0121 [hep-ph]}}.

\bibitem{Dumitru:2008wn}
A.~Dumitru, F.~Gelis, L.~McLerran, and R.~Venugopalan, ``{Glasma flux tubes and
  the near side ridge phenomenon at RHIC}'',
  \href{http://dx.doi.org/10.1016/j.nuclphysa.2008.06.012}{{\em Nucl. Phys. A}
  {\bfseries 810} (2008) 91}, \href{http://arxiv.org/abs/0804.3858}{{\ttfamily
  arXiv:0804.3858 [hep-ph]}}.

\bibitem{Andersson:1983ia}
B.~Andersson, G.~Gustafson, G.~Ingelman, and T.~Sjostrand, ``{Parton
  Fragmentation and String Dynamics}'',
  \href{http://dx.doi.org/10.1016/0370-1573(83)90080-7}{{\em Phys. Rept.}
  {\bfseries 97} (1983) 31--145}.

\bibitem{Braun-Munzinger:2018yru}
P.~Braun-Munzinger, A.~Rustamov, and J.~Stachel, ``{Experimental results on
  fluctuations of conserved charges confronted with predictions from canonical
  thermodynamics}'',
  \href{http://dx.doi.org/10.1016/j.nuclphysa.2018.09.074}{{\em Nucl. Phys. A}
  {\bfseries 982} (2019) 307},
  \href{http://arxiv.org/abs/1807.08927}{{\ttfamily arXiv:1807.08927
  [nucl-th]}}.

\bibitem{HotQCD:2017qwq}
{\bfseries HotQCD} Collaboration, A.~Bazavov {\em et~al.}, ``{Skewness and
  kurtosis of net baryon-number distributions at small values of the baryon
  chemical potential}'',
  \href{http://dx.doi.org/10.1103/PhysRevD.96.074510}{{\em Phys. Rev. D}
  {\bfseries 96} (2017) 074510},
  \href{http://arxiv.org/abs/1708.04897}{{\ttfamily arXiv:1708.04897
  [hep-lat]}}.

\bibitem{Citron:2018lsq}
Z.~Citron {\em et~al.}, ``{Report from Working Group 5}: {Future physics
  opportunities for high-density QCD at the LHC with heavy-ion and proton
  beams}'', \href{http://dx.doi.org/10.23731/CYRM-2019-007.1159}{{\em CERN
  Yellow Rep. Monogr.} {\bfseries 7} (2019) 1159},
  \href{http://arxiv.org/abs/1812.06772}{{\ttfamily arXiv:1812.06772
  [hep-ph]}}.

\bibitem{ALICE:2803563}
{\bfseries ALICE} Collaboration, ``{Letter of intent for ALICE 3: A next
  generation heavy-ion experiment at the LHC}'', {\em CERN-LHCC-2022-009,
  LHCC-I-038} (2022) . \url{http://cds.cern.ch/record/2803563}.

\end{thebibliography}\endgroup

\newpage
\appendix

%
%

\section{The ALICE Collaboration}
\label{app:collab}
\begin{flushleft} 
\small

S.~Acharya\,\orcidlink{0000-0002-9213-5329}\,$^{\rm 125,132}$, 
D.~Adamov\'{a}\,\orcidlink{0000-0002-0504-7428}\,$^{\rm 86}$, 
A.~Adler$^{\rm 69}$, 
G.~Aglieri Rinella\,\orcidlink{0000-0002-9611-3696}\,$^{\rm 32}$, 
M.~Agnello\,\orcidlink{0000-0002-0760-5075}\,$^{\rm 29}$, 
N.~Agrawal\,\orcidlink{0000-0003-0348-9836}\,$^{\rm 50}$, 
Z.~Ahammed\,\orcidlink{0000-0001-5241-7412}\,$^{\rm 132}$, 
S.~Ahmad\,\orcidlink{0000-0003-0497-5705}\,$^{\rm 15}$, 
S.U.~Ahn\,\orcidlink{0000-0001-8847-489X}\,$^{\rm 70}$, 
I.~Ahuja\,\orcidlink{0000-0002-4417-1392}\,$^{\rm 37}$
% ...

\section*{Affiliation Notes}

$^{\rm I}$ Deceased\\
$^{\rm II}$ Also at: Max-Planck-Institut f\"{u}r Physik, Munich, Germany\\
$^{\rm III}$ Also at: Italian National Agency for New Technologies, Energy and Sustainable Economic Development (ENEA), Bologna, Italy\\
$^{\rm IV}$ Also at: Dipartimento DET del Politecnico di Torino, Turin, Italy\\
$^{\rm V}$ Also at: Department of Applied Physics, Aligarh Muslim University, Aligarh, India\\
$^{\rm VI}$ Also at: Institute of Theoretical Physics, University of Wroclaw, Poland\\
$^{\rm VII}$ Also at: An institution covered by a cooperation agreement with CERN\\

\section*{Collaboration Institutes}

$^{1}$ A.I. Alikhanyan National Science Laboratory (Yerevan Physics Institute) Foundation, Yerevan, Armenia\\
$^{2}$ AGH University of Science and Technology, Cracow, Poland\\
$^{3}$ Bogolyubov Institute for Theoretical Physics, National Academy of Sciences of Ukraine, Kiev, Ukraine\\
$^{4}$ Bose Institute, Department of Physics  and Centre for Astroparticle Physics and Space Science (CAPSS), Kolkata, India\\
$^{5}$ California Polytechnic State University, San Luis Obispo, California, United States\\
$^{6}$ Central China Normal University, Wuhan, China\\
$^{7}$ Centro de Aplicaciones Tecnol\'{o}gicas y Desarrollo Nuclear (CEADEN), Havana, Cuba\\
$^{8}$ Centro de Investigaci\'{o}n y de Estudios Avanzados (CINVESTAV), Mexico City and M\'{e}rida, Mexico\\
$^{9}$ Chicago State University, Chicago, Illinois, United States\\
$^{10}$ China Institute of Atomic Energy, Beijing, China\\
$^{11}$ Chungbuk National University, Cheongju, Republic of Korea\\
$^{12}$ Comenius University Bratislava, Faculty of Mathematics, Physics and Informatics, Bratislava, Slovak Republic\\
$^{13}$ COMSATS University Islamabad, Islamabad, Pakistan\\
$^{14}$ Creighton University, Omaha, Nebraska, United States\\
$^{15}$ Department of Physics, Aligarh Muslim University, Aligarh, India\\
$^{16}$ Department of Physics, Pusan National University, Pusan, Republic of Korea\\
$^{17}$ Department of Physics, Sejong University, Seoul, Republic of Korea\\
$^{18}$ Department of Physics, University of California, Berkeley, California, United States\\
$^{19}$ Department of Physics, University of Oslo, Oslo, Norway\\
$^{20}$ Department of Physics and Technology, University of Bergen, Bergen, Norway\\
$^{21}$ Dipartimento di Fisica, Universit\`{a} di Pavia, Pavia, Italy\\
$^{22}$ Dipartimento di Fisica dell'Universit\`{a} and Sezione INFN, Cagliari, Italy\\
$^{23}$ Dipartimento di Fisica dell'Universit\`{a} and Sezione INFN, Trieste, Italy\\
$^{24}$ Dipartimento di Fisica dell'Universit\`{a} and Sezione INFN, Turin, Italy\\
$^{25}$ Dipartimento di Fisica e Astronomia dell'Universit\`{a} and Sezione INFN, Bologna, Italy\\
$^{26}$ Dipartimento di Fisica e Astronomia dell'Universit\`{a} and Sezione INFN, Catania, Italy\\
$^{27}$ Dipartimento di Fisica e Astronomia dell'Universit\`{a} and Sezione INFN, Padova, Italy\\
$^{28}$ Dipartimento di Fisica `E.R.~Caianiello' dell'Universit\`{a} and Gruppo Collegato INFN, Salerno, Italy\\
$^{29}$ Dipartimento DISAT del Politecnico and Sezione INFN, Turin, Italy\\
$^{30}$ Dipartimento di Scienze MIFT, Universit\`{a} di Messina, Messina, Italy\\
$^{31}$ Dipartimento Interateneo di Fisica `M.~Merlin' and Sezione INFN, Bari, Italy\\
$^{32}$ European Organization for Nuclear Research (CERN), Geneva, Switzerland\\
$^{33}$ Faculty of Electrical Engineering, Mechanical Engineering and Naval Architecture, University of Split, Split, Croatia\\
$^{34}$ Faculty of Engineering and Science, Western Norway University of Applied Sciences, Bergen, Norway\\
$^{35}$ Faculty of Nuclear Sciences and Physical Engineering, Czech Technical University in Prague, Prague, Czech Republic\\
$^{36}$ Faculty of Physics, Sofia University, Sofia, Bulgaria\\
$^{37}$ Faculty of Science, P.J.~\v{S}af\'{a}rik University, Ko\v{s}ice, Slovak Republic\\
$^{38}$ Frankfurt Institute for Advanced Studies, Johann Wolfgang Goethe-Universit\"{a}t Frankfurt, Frankfurt, Germany\\
$^{39}$ Fudan University, Shanghai, China\\
$^{40}$ Gangneung-Wonju National University, Gangneung, Republic of Korea\\
$^{41}$ Gauhati University, Department of Physics, Guwahati, India\\
$^{42}$ Helmholtz-Institut f\"{u}r Strahlen- und Kernphysik, Rheinische Friedrich-Wilhelms-Universit\"{a}t Bonn, Bonn, Germany\\
$^{43}$ Helsinki Institute of Physics (HIP), Helsinki, Finland\\
$^{44}$ High Energy Physics Group,  Universidad Aut\'{o}noma de Puebla, Puebla, Mexico\\
$^{45}$ Horia Hulubei National Institute of Physics and Nuclear Engineering, Bucharest, Romania\\
$^{46}$ Indian Institute of Technology Bombay (IIT), Mumbai, India\\
$^{47}$ Indian Institute of Technology Indore, Indore, India\\
$^{48}$ INFN, Laboratori Nazionali di Frascati, Frascati, Italy\\
$^{49}$ INFN, Sezione di Bari, Bari, Italy\\
$^{50}$ INFN, Sezione di Bologna, Bologna, Italy\\
$^{51}$ INFN, Sezione di Cagliari, Cagliari, Italy\\
$^{52}$ INFN, Sezione di Catania, Catania, Italy\\
$^{53}$ INFN, Sezione di Padova, Padova, Italy\\
$^{54}$ INFN, Sezione di Pavia, Pavia, Italy\\
$^{55}$ INFN, Sezione di Torino, Turin, Italy\\
$^{56}$ INFN, Sezione di Trieste, Trieste, Italy\\
$^{57}$ Inha University, Incheon, Republic of Korea\\
$^{58}$ Institute for Gravitational and Subatomic Physics (GRASP), Utrecht University/Nikhef, Utrecht, Netherlands\\
$^{59}$ Institute of Experimental Physics, Slovak Academy of Sciences, Ko\v{s}ice, Slovak Republic\\
$^{60}$ Institute of Physics, Homi Bhabha National Institute, Bhubaneswar, India\\
$^{61}$ Institute of Physics of the Czech Academy of Sciences, Prague, Czech Republic\\
$^{62}$ Institute of Space Science (ISS), Bucharest, Romania\\
$^{63}$ Institut f\"{u}r Kernphysik, Johann Wolfgang Goethe-Universit\"{a}t Frankfurt, Frankfurt, Germany\\
$^{64}$ Instituto de Ciencias Nucleares, Universidad Nacional Aut\'{o}noma de M\'{e}xico, Mexico City, Mexico\\
$^{65}$ Instituto de F\'{i}sica, Universidade Federal do Rio Grande do Sul (UFRGS), Porto Alegre, Brazil\\
$^{66}$ Instituto de F\'{\i}sica, Universidad Nacional Aut\'{o}noma de M\'{e}xico, Mexico City, Mexico\\
$^{67}$ iThemba LABS, National Research Foundation, Somerset West, South Africa\\
$^{68}$ Jeonbuk National University, Jeonju, Republic of Korea\\
$^{69}$ Johann-Wolfgang-Goethe Universit\"{a}t Frankfurt Institut f\"{u}r Informatik, Fachbereich Informatik und Mathematik, Frankfurt, Germany\\
$^{70}$ Korea Institute of Science and Technology Information, Daejeon, Republic of Korea\\
$^{71}$ KTO Karatay University, Konya, Turkey\\
$^{72}$ Laboratoire de Physique des 2 Infinis, Ir\`{e}ne Joliot-Curie, Orsay, France\\
$^{73}$ Laboratoire de Physique Subatomique et de Cosmologie, Universit\'{e} Grenoble-Alpes, CNRS-IN2P3, Grenoble, France\\
$^{74}$ Lawrence Berkeley National Laboratory, Berkeley, California, United States\\
$^{75}$ Lund University Department of Physics, Division of Particle Physics, Lund, Sweden\\
$^{76}$ Nagasaki Institute of Applied Science, Nagasaki, Japan\\
$^{77}$ Nara Women{'}s University (NWU), Nara, Japan\\
$^{78}$ National and Kapodistrian University of Athens, School of Science, Department of Physics , Athens, Greece\\
$^{79}$ National Centre for Nuclear Research, Warsaw, Poland\\
$^{80}$ National Institute of Science Education and Research, Homi Bhabha National Institute, Jatni, India\\
$^{81}$ National Nuclear Research Center, Baku, Azerbaijan\\
$^{82}$ National Research and Innovation Agency - BRIN, Jakarta, Indonesia\\
$^{83}$ Niels Bohr Institute, University of Copenhagen, Copenhagen, Denmark\\
$^{84}$ Nikhef, National institute for subatomic physics, Amsterdam, Netherlands\\
$^{85}$ Nuclear Physics Group, STFC Daresbury Laboratory, Daresbury, United Kingdom\\
$^{86}$ Nuclear Physics Institute of the Czech Academy of Sciences, Husinec-\v{R}e\v{z}, Czech Republic\\
$^{87}$ Oak Ridge National Laboratory, Oak Ridge, Tennessee, United States\\
$^{88}$ Ohio State University, Columbus, Ohio, United States\\
$^{89}$ Physics department, Faculty of science, University of Zagreb, Zagreb, Croatia\\
$^{90}$ Physics Department, Panjab University, Chandigarh, India\\
$^{91}$ Physics Department, University of Jammu, Jammu, India\\
$^{92}$ Physics Department, University of Rajasthan, Jaipur, India\\
$^{93}$ Physics Program and International Institute for Sustainability with Knotted Chiral Meta Matter (SKCM2), Hiroshima University, Hiroshima, Japan\\
$^{94}$ Physikalisches Institut, Eberhard-Karls-Universit\"{a}t T\"{u}bingen, T\"{u}bingen, Germany\\
$^{95}$ Physikalisches Institut, Ruprecht-Karls-Universit\"{a}t Heidelberg, Heidelberg, Germany\\
$^{96}$ Physik Department, Technische Universit\"{a}t M\"{u}nchen, Munich, Germany\\
$^{97}$ Politecnico di Bari and Sezione INFN, Bari, Italy\\
$^{98}$ Research Division and ExtreMe Matter Institute EMMI, GSI Helmholtzzentrum f\"ur Schwerionenforschung GmbH, Darmstadt, Germany\\
$^{99}$ Saga University, Saga, Japan\\
$^{100}$ Saha Institute of Nuclear Physics, Homi Bhabha National Institute, Kolkata, India\\
$^{101}$ School of Physics and Astronomy, University of Birmingham, Birmingham, United Kingdom\\
$^{102}$ Secci\'{o}n F\'{\i}sica, Departamento de Ciencias, Pontificia Universidad Cat\'{o}lica del Per\'{u}, Lima, Peru\\
$^{103}$ Stefan Meyer Institut f\"{u}r Subatomare Physik (SMI), Vienna, Austria\\
$^{104}$ SUBATECH, IMT Atlantique, Nantes Universit\'{e}, CNRS-IN2P3, Nantes, France\\
$^{105}$ Suranaree University of Technology, Nakhon Ratchasima, Thailand\\
$^{106}$ Technical University of Ko\v{s}ice, Ko\v{s}ice, Slovak Republic\\
$^{107}$ The Henryk Niewodniczanski Institute of Nuclear Physics, Polish Academy of Sciences, Cracow, Poland\\
$^{108}$ The University of Texas at Austin, Austin, Texas, United States\\
$^{109}$ Universidad Aut\'{o}noma de Sinaloa, Culiac\'{a}n, Mexico\\
$^{110}$ Universidade de S\~{a}o Paulo (USP), S\~{a}o Paulo, Brazil\\
$^{111}$ Universidade Estadual de Campinas (UNICAMP), Campinas, Brazil\\
$^{112}$ Universidade Federal do ABC, Santo Andre, Brazil\\
$^{113}$ University of Cape Town, Cape Town, South Africa\\
$^{114}$ University of Houston, Houston, Texas, United States\\
$^{115}$ University of Jyv\"{a}skyl\"{a}, Jyv\"{a}skyl\"{a}, Finland\\
$^{116}$ University of Kansas, Lawrence, Kansas, United States\\
$^{117}$ University of Liverpool, Liverpool, United Kingdom\\
$^{118}$ University of Science and Technology of China, Hefei, China\\
$^{119}$ University of South-Eastern Norway, Kongsberg, Norway\\
$^{120}$ University of Tennessee, Knoxville, Tennessee, United States\\
$^{121}$ University of the Witwatersrand, Johannesburg, South Africa\\
$^{122}$ University of Tokyo, Tokyo, Japan\\
$^{123}$ University of Tsukuba, Tsukuba, Japan\\
$^{124}$ University Politehnica of Bucharest, Bucharest, Romania\\
$^{125}$ Universit\'{e} Clermont Auvergne, CNRS/IN2P3, LPC, Clermont-Ferrand, France\\
$^{126}$ Universit\'{e} de Lyon, CNRS/IN2P3, Institut de Physique des 2 Infinis de Lyon, Lyon, France\\
$^{127}$ Universit\'{e} de Strasbourg, CNRS, IPHC UMR 7178, F-67000 Strasbourg, France, Strasbourg, France\\
$^{128}$ Universit\'{e} Paris-Saclay Centre d'Etudes de Saclay (CEA), IRFU, D\'{e}partment de Physique Nucl\'{e}aire (DPhN), Saclay, France\\
$^{129}$ Universit\`{a} degli Studi di Foggia, Foggia, Italy\\
$^{130}$ Universit\`{a} del Piemonte Orientale, Vercelli, Italy\\
$^{131}$ Universit\`{a} di Brescia, Brescia, Italy\\
$^{132}$ Variable Energy Cyclotron Centre, Homi Bhabha National Institute, Kolkata, India\\
$^{133}$ Warsaw University of Technology, Warsaw, Poland\\
$^{134}$ Wayne State University, Detroit, Michigan, United States\\
$^{135}$ Westf\"{a}lische Wilhelms-Universit\"{a}t M\"{u}nster, Institut f\"{u}r Kernphysik, M\"{u}nster, Germany\\
$^{136}$ Wigner Research Centre for Physics, Budapest, Hungary\\
$^{137}$ Yale University, New Haven, Connecticut, United States\\
$^{138}$ Yonsei University, Seoul, Republic of Korea\\
$^{139}$  Zentrum  f\"{u}r Technologie und Transfer (ZTT), Worms, Germany\\
$^{140}$ Affiliated with an institute covered by a cooperation agreement with CERN\\
$^{141}$ Affiliated with an international laboratory covered by a cooperation agreement with CERN.\\

\end{flushleft} 

\end{document}